\documentclass{emulateapj}

\usepackage{xspace}
\usepackage{graphicx}
\usepackage{natbib}
\usepackage{apjfonts}
\newcommand\suzaku{\textsl{Suzaku}\xspace}
\newcommand\bsax{\textsl{BeppoSAX}\xspace}

\newcommand\swift{\textsl{Swift}\xspace}
\newcommand\osse{\textsl{OSSE}\xspace}

\newcommand\hxd{\textsl{HXD}\xspace}

\newcommand\chandra{\textsl{Chandra}\xspace}
\newcommand\xmm{\textsl{XMM-Newton}\xspace}

\newcommand\integral{\textsl{INTEGRAL}\xspace}

\newcommand\asca{\textsl{ASCA}\xspace}

\newcommand\pexmon{\texttt{PEXMON}\xspace}
\newcommand\hetg{{HETG}\xspace}

\newcommand\xis{{XIS}\xspace}
\newcommand\pin{{PIN}\xspace}
\newcommand\heg{{HEG}\xspace}
\newcommand\meg{{MEG}\xspace}
\newcommand\acis{{ACIS}\xspace}

\newcommand\myt{{\tt MYTorus}\xspace}

\newcommand\fek{{Fe~K}\xspace}
\newcommand\feka{{Fe~K$\alpha$}\xspace}
\newcommand\fekb{{Fe~K$\beta$}\xspace}
\newcommand\nika{{Ni~K$\alpha$}\xspace}

\newcommand\fexxv{Fe~{\sc xxv}\xspace}
\newcommand\fexxvi{Fe~{\sc xxvi}\xspace}

\newcommand\nh{\textsl{$N_{\rm H}$}\xspace}

\newcommand\mcgeight{{MCG~+8-11-11}\xspace}

\newcommand\aproxgt{\mathrel{%
      \rlap{\raise 0.511ex \hbox{$>$}}{\lower 0.511ex \hbox{$\sim$}}}}
\newcommand\aproxlt{\mathrel{%
      \rlap{\raise 0.511ex \hbox{$<$}}{\lower 0.511ex \hbox{$\sim$}}}}
      
\slugcomment{Submitted 2014 April 11}
\shorttitle{A \chandra-\hetg view of \mcgeight}  
\shortauthors{Murphy \& Nowak}

\begin{document}

\title {A \chandra-\hetg view of \mcgeight}  

\author{K. D. Murphy}
\affil{Department of Physics, Skidmore College, Saratoga Springs, NY 12866}
\email{kmurphy1@skidmore.edu}

\and

\author{M. A. Nowak} 
\affil{MIT Kavli Institute for Space Research, 77 Massachusetts Ave.,
  NE83-653, Cambridge, MA 02139}

\begin{abstract}
We present a spectral analysis of the 118\,ks High Energy Transmission
Gratings (\hetg) observation of the X-ray bright Seyfert~1.5 galaxy
\mcgeight, in conjunction with 100\,ks of archival \suzaku data, aimed
at investigating the signatures of warm absorption and Compton
reflection reported from previous \suzaku and \xmm studies of the
source.  Contrary to previous results, we find that warm absorption is
not required by the data. Instead, we report upper limits on
absorption lines that are below previous (marginal) detections.  \feka
line emission is clearly detected and is likely resolved with $\sigma
\sim 0.02$\,keV with the \hetg data.  We applied self-consistent,
broadband spectral-fitting models to the \hetg and \suzaku data to
investigate this and other signatures of distant absorption and
reflection.  Utilizing in particular the \myt model, we find that the
data are consistent with reprocessing by a distant, neutral torus that
is Compton thick (\nh$\sim 10^{24} \rm cm^{-2}$) and out of the
line-of-sight.  However, we do not find compelling evidence of a
relativistically-broadened Fe-K emission line, which is often expected
from type~1 AGN. This is consistent with some, although not all,
previous studies of \mcgeight.  A well-measured edge is identified by
the \hetg near 0.5\,keV, indicating neutral absorption in the line of
sight that is consistent with galactic absorption; however, the
absorption may be partially intrinsic to the source.  The \hetg data
are consistent with the presence of a soft excess, a feature that may
be missed by considering the \suzaku data alone.
\end{abstract}


\section{Introduction} 
\label{sec:intro}

The unified model of Active Galactic Nuclei (AGN) postulates a system
of complex components: an inner accretion disk near the black hole
responsible for the bulk of the bolometric emission, outer Broad and
Narrow Line Regions (BLR/NLR), and a putative torus also at outer
radii, possibly obscuring the inner emission regions for some viewing
angles.  Recent X-ray studies have attempted to use the inner
accretion flows of these systems to probe the relativistic nature of
the black hole (i.e., by measuring relativistically broadened \feka
lines; see \citealt{reynolds:03a} for a review).  The interpretation
of a broadened line has come under criticism, however, from some
authors who point out that the 2--10\,keV continuum, and specifically
the Fe line and edge region, might be sculpted by the presence of a
warm, highly ionized absorber/wind \citep{miller:09a}.  Understanding
the nature of the highly ionized, absorbing wind (whose presence has
been recognized since CCD spectroscopic observations with the Advanced
Satellite for for Cosmology and Astrophysics, \asca;
\citealt{reynolds:95a}) and cold reflection is therefore crucial for
both developing global models of AGN systems, as well as for
accurately describing relativistic emission from the inner accretion
disk.

\mcgeight, a nearby ($z=0.0205$) Seyfert~1.5 galaxy, is one of the
X-ray brightest Seyfert galaxies ever observed ($L_{2-10}\sim 10^{44}
\ \rm erg \ s^{-1}$).  The broadband X-ray spectrum has historically
been described by a power law with a high-energy, exponential cutoff,
a strong \feka emission line, a Compton reflection component, and warm
absorption (\citealt{grandi:98a}; \citealt{perola:00a}); its X-ray
spectral shape has not been seen to vary significantly since it was
observed by \asca (\citealt{bianchi:10a}).  A study of this source
with \xmm \citep{matt:06a} was conducted to determine the origin of
the Fe~K line emission.  Evidence of a Compton shoulder associated
with the unresolved \feka line emission was found, suggesting,
together with the clear reflection component in the continuum, that
the emission originated in distant matter that may be Compton thick.
Only a loose upper limit on the ratio of the Compton shoulder, which
was modeled ad-hoc with Gaussian line emission, to the \feka line core
emission was determined.  Marginal detections of \fekb and \nika,
likely originating in the same structure, were also reported, but only
loose constraints on the ratios of these lines were possible.  One of
the more puzzling results of this study was the lack of detection of
an underlying broad component (i.e., from the accretion disk) of the
\feka emission line, which is expected to be seen in the X-ray
spectrum of bright type 1 AGN, as their inclination angles
theoretically afford us an unobscured view of their innermost regions.
However, some ambiguity was found: the equivalent width of the \feka
line core plus Compton shoulder was found to be $\sim 90$\,eV, implying
either an underabundance of iron or the existence of two reflection
regions (i.e., a torus and an ionized accretion disk, the latter
providing a reflection hump but not a narrow \fek emission line;
\citealt{matt:06a}).

\begin{figure*}
\begin{center}
\includegraphics[width=0.47\textwidth,bb=90 25 597 379]{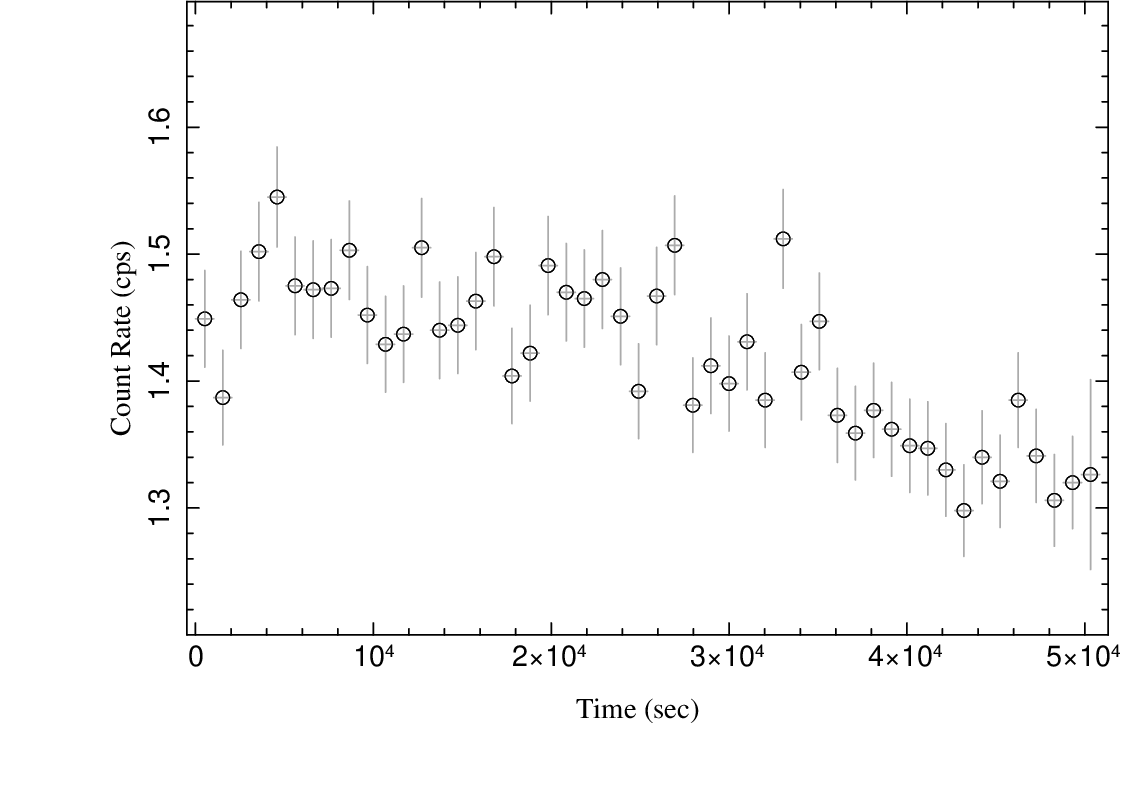}
\includegraphics[width=0.47\textwidth,bb=90 25 597 379]{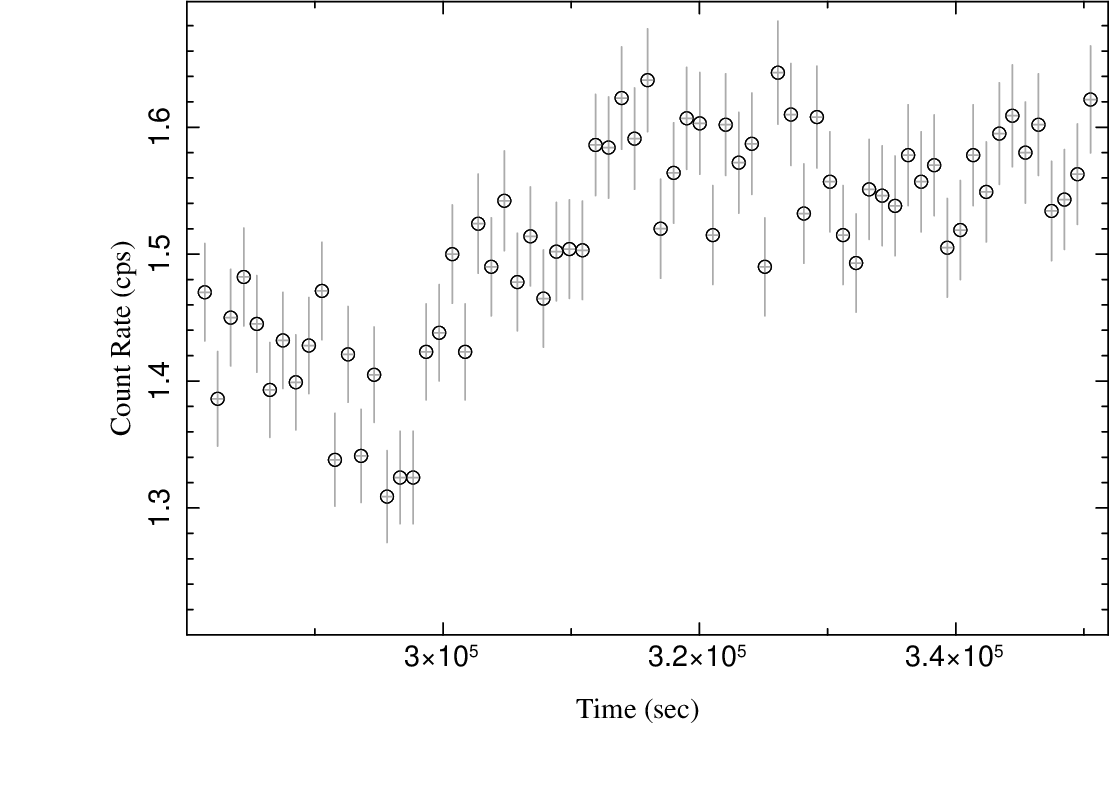}
\end{center}
\caption{\chandra-\hetg light curves in 1000\,s bins from the combined
  $\pm1^{\rm st}$ orders of the \heg and \meg in the 1.7--25\,\AA\
  band.
\label{fig:ltcrv}}
\end{figure*}
 
The \xmm data \citep{matt:06a} also required (at CCD resolution) the
inclusion of a highly ionized warm absorber ($\log \xi\approx 2.7$,
$\log N_{\rm H} \approx 23$), as well as an edge in the $< 1$\ keV
spectrum.  A number of possible absorption lines were found in the
Reflection Gratings Spectrometer (RGS) spectrum, but their
identification was ambiguous.  However, such a highly ionized absorber
would have placed most ionized absorption lines at energies outside of
the useful range of the RGS detector. Although the velocity structure
could not be deduced from the absorption edges, the velocity structure
appeared to be complex.

\citet{bianchi:10a} and \citet{patrick:12a} found results consistent
with those reported by \citet{matt:06a} in their analyses of the
100\,ks \suzaku X-ray Imaging Spectrometer (\xis) observation of
\mcgeight, namely an intrinsic power-law spectrum with a photon index
of $\Gamma=1.7$--1.8, a lack of soft excess, warm absorption in the
line-of-sight (required by spectral curvature at low energies), and
significant Compton reflection.  Both reported evidence of \fexxvi
line emission; \citet{bianchi:10a} also reported a detection of \fexxv
line emission.  However, unlike \citet{matt:06a}, \citet{patrick:12a} found
that the data require relativistic \feka line emission in addition to
narrow line emission.  \citet{bianchi:10a} did not find evidence of
this component when considering \xis data alone; however, it became
required when the Hard X-ray Detector (\hxd) \pin data were included
in their analysis.  In this case, the reflection fraction became
significantly smaller, the (weak) reflection continuum was associated
with the accretion disk, and the origin of the narrow Fe~K emission
was attributed to a Compton-thin region (the BLR).  An analysis of
\suzaku data together with \integral and \swift data by
\citet{soldi:10a} also pointed to little or no Compton reflection.
 Both \citet{bianchi:10a} and \citet{soldi:10a} quantified the amount
of reflection with the reflection fraction obtained by fitting the {\tt PEXRAV}
model to the broadband data; they found values of $R=0.2-0.3$ and $R<0.2$, 
respectively, in contrast with $R\sim1$ found by \citet{bianchi:10a} with
the \xis data alone.  \citet{bianchi:10a} furthermore found that the broadband spectrum
indicated sub-solar Fe abundance, but noted that there is ambiguity between the
iron abundance and the inclination angle of the system.  Additional
ambiguity was found between the broad \feka line emission and the
possible Compton shoulder redward of the narrow \feka line emission.

We obtained $\approx 118$\,ks of \chandra High Energy Transmission
Gratings (\hetg) data for \mcgeight to investigate the presence of
absorption lines from a potentially multi-zoned warm absorber as well
as the structure and location of the origin of the narrow \feka line
emission by disentangling it from the Fe K line complex.  As pointed
out by, e.g., \citet{bianchi:10a}, the analysis of the X-ray spectrum
may yield compromised results if the covered energy band is limited to
below 10\,keV.  We therefore report on our analysis of the \chandra
data in combination with the archived, broadband data from \suzaku.

\section{Observations and Data Reduction}
\subsection{\chandra-\hetg}
Observations of \mcgeight were completed on 2010 December 7 (obsid
12861) and 2010 December 10 (obsid 13200) by the \chandra Advanced CCD
Imaging Spectrometer (\acis; \citealt{garmire:03a}), using the High
Energy Transmission Gratings (\hetg; \citealt{canizares:05a}).
Exposure times of 49.2\,ks and 68.9\,ks, respectively, were obtained.
Light curves for the two observations are shown in
Fig.~\ref{fig:ltcrv}, which exhibit little variability over the course
of the two observations.
 
We used CIAO version 4.4 and CALDB version 4.5.1 to run a standard
extraction of the first-order High Energy Gratings (\heg) and Medium
Energy Gratings (\meg) data and to create the response files.  We
first investigated each of the \heg and \meg data sets for the two
observations separately.  Finding good agreement among the spectra,
and in order to achieve the highest signal-to-noise ratio possible, we
co-added the data (using the \texttt{combine\_datasets} function in
the \texttt{Interactive Spectral Interpretation System} --
\texttt{ISIS}; \citealt{houck:00a}) from the two observations and from
the positive and negative first orders for each of the transmission
gratings to obtain a single \heg and a single \meg data set.  To
preserve the higher spectral resolution of the \heg in the Fe~K band,
we did not co-add the \heg and \meg data for the following analysis.
 
\subsection{\suzaku-\xis and \hxd-\pin}

\begin{figure*}
\begin{center}
\includegraphics[width=0.48\textwidth,bb=90 25 597 379]{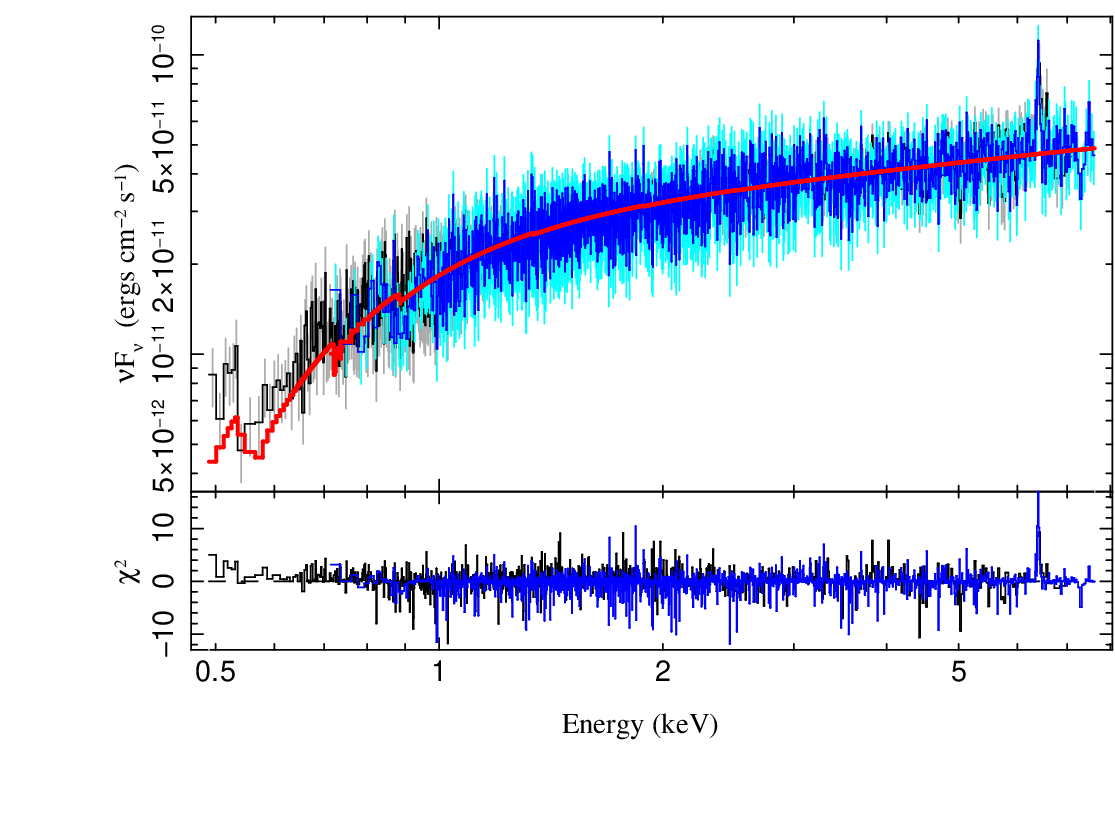}
\includegraphics[width=0.48\textwidth,bb=90 25 597 379]{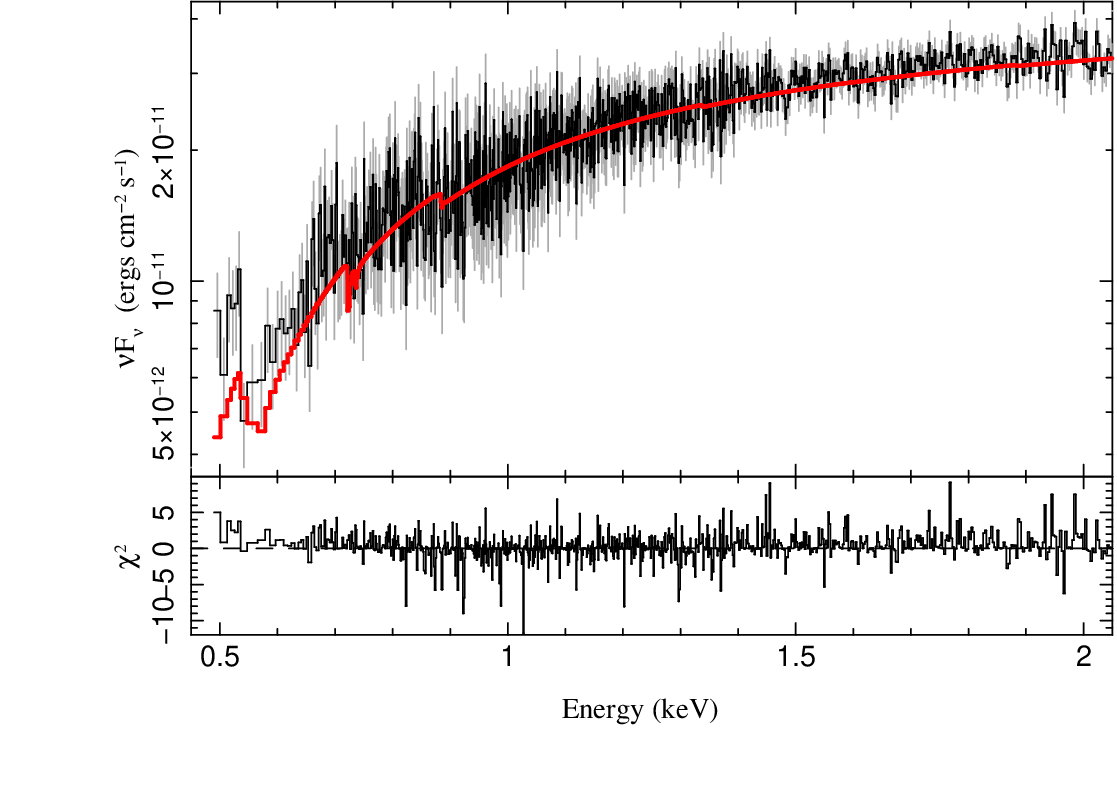}
\includegraphics[width=0.48\textwidth,bb=90 25 597 379]{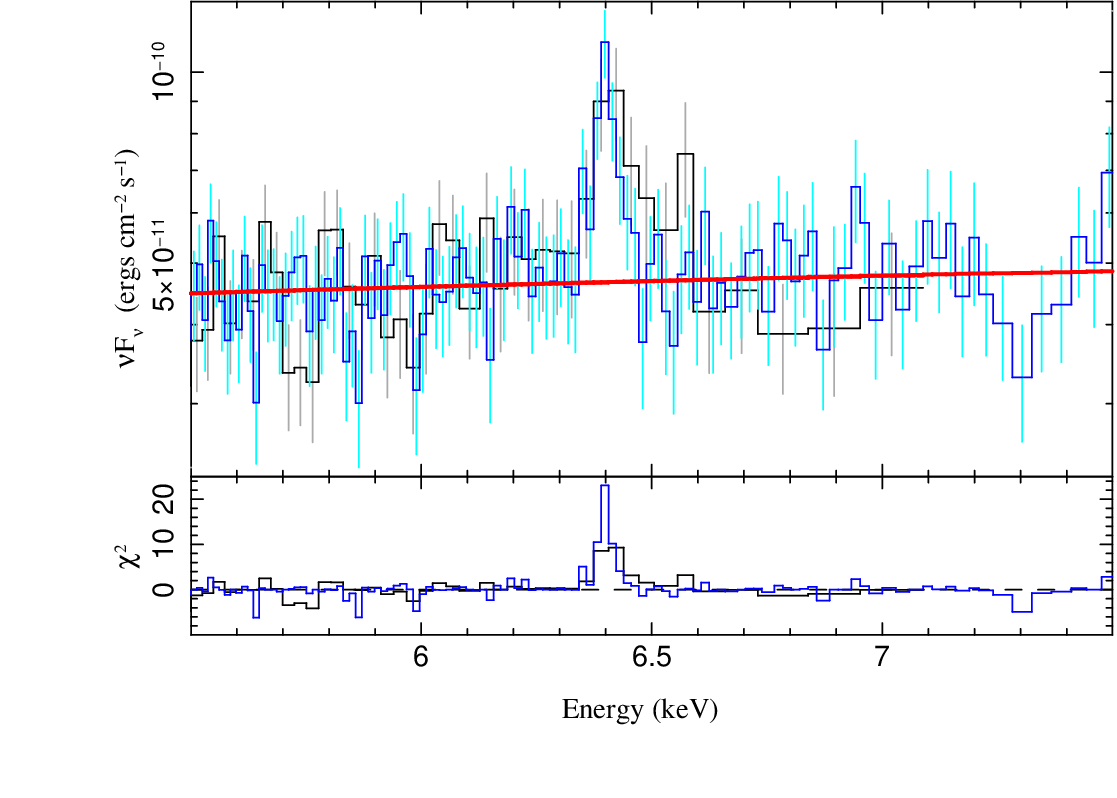}
\includegraphics[width=0.48\textwidth,bb=90 25 597 379]{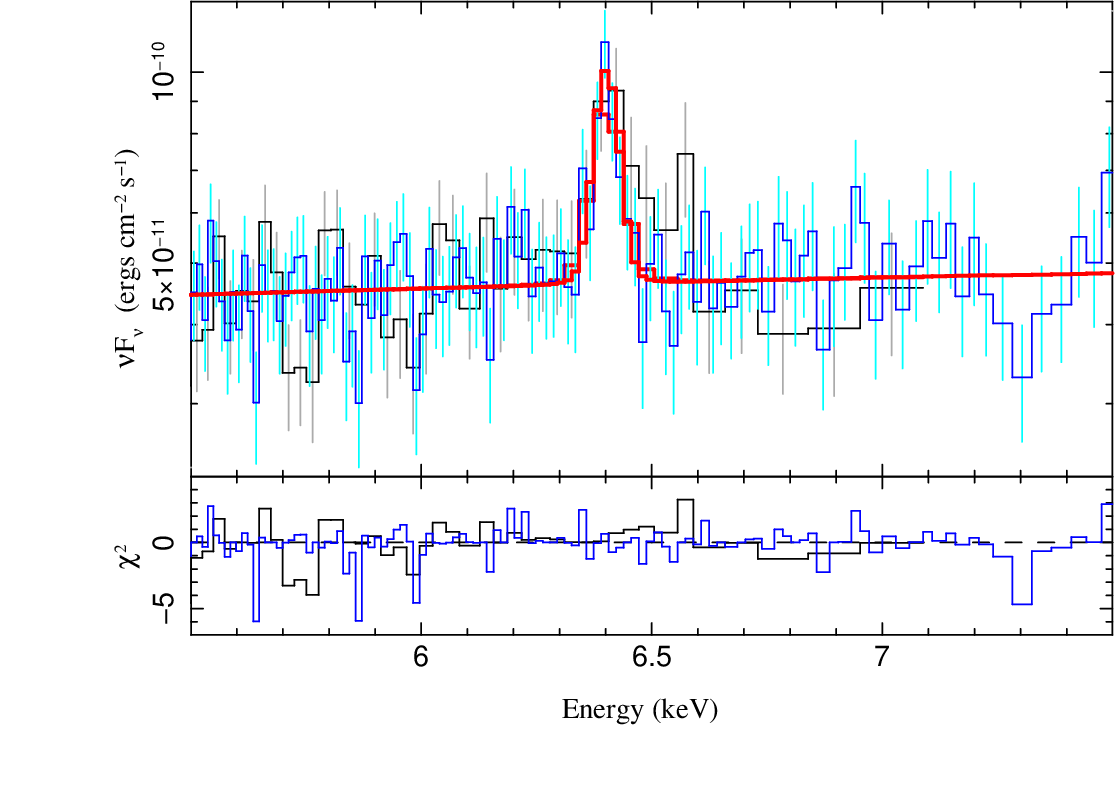}
\end{center}
\caption{{\it Top left and right, bottom left:} A simple absorbed power law model
(red) was applied to \chandra-\meg (black) and \heg (blue) spectra.
For this and subsequent figures, colors appear only in the 
online version of the paper.
The figure on the top left shows the full flux-corrected spectra.  Close-ups
of the soft band (\meg only) and \fek band are shown in the top right and bottom
left panels, respectively; clear residuals are
seen in the low-energy \meg
spectrum and near the rest energy of \feka.  {\it Bottom right:} The model and flux-corrected spectra in the \fek
band are shown after a gaussian line is added to the model.
\label{fig:initmodelpl}}
\end{figure*}

\mcgeight was observed by \suzaku on 2007 September 17 for 100\,ks
(obsid 702112010), using the \hxd nominal
pointing.  We followed a similar method of reducing the \suzaku data
as described in detail by \citet{bianchi:10a}, using the June 6, 2011
release of the \suzaku calibration files and \texttt{HEASOFT} version
6.11.  For the observation with the \xis CCDs, we chose circular
extraction regions with approximately 4.5\,arcmin radii. For the \xis
background we extracted spectra from source-free, rectangular regions
(approximately 12\,arcmin $\times$ 4\,arcmin) near the edge of the
chips.  Individual spectra were extracted for each \xis CCD detector
(i.e., \xis0, \xis1, and \xis3) and separate data acquisition mode
($3\times3$ and $5\times5$ mode), and then co-added during spectral
analysis using the \texttt{ISIS} \texttt{combine\_datasets} function.
For the \hxd we used only the \pin spectra, and
adopted the Cosmic X-ray Background model recommended by the \suzaku
ABC guide.  The recommended cross-normalization of 1.18 was used
between the \pin and \xis data throughout our analysis.

\section{Data Analysis \& Results}

We analyzed the \chandra and \suzaku data using \texttt{ISIS} version
1.6.2.  For both the \heg and \meg spectra, we grouped the data to a
minimum signal-to-noise of 4.5 and two energy channels per bin, and
considered data in the 0.7--7.5\,keV and 0.5--7.0\,keV range,
respectively. For the \suzaku-\xis spectra we grouped the data to a
minimum signal-to-noise of 6 and a minimum channel criterion that
ensured that the binning was no finer than the half width half maximum
resolution of the detector (see the description of this channel
binning in \citealt{nowak:11a}).  For the \xis spectra we considered
data in the 0.8--1.7\,keV and 2.4--9\,keV range.  For the \hxd-\pin
data we also binned to a minimum signal-to-noise of 6, and considered
data in the 15--60\,keV range.  In all of our spectral fits, we used
the \citet{wilms:00a} abundances and the photoelectric absorption
cross-sections of \citet{verner:96a}.  The assumed cosmological
parameters are $H_0=73 \rm \ km \ s^{-1} \ Mpc^{-1}$,
$\Omega_{\Lambda}$=0.7, $\Omega_m$=0.3.  Unless otherwise noted, all
spectral plots are in the rest frame of \mcgeight.  We quote
statistical errors corresponding to 90\% confidence for one
interesting parameter ($\Delta\chi^2=2.71$).

\subsection{Preliminary Analysis}

\label{sec:phenom}

To guide us in the adoption of an applicable global model, we first
investigated the data with a phenomenological model, adding components
as necessary to improve the fit.  Here we outline the components
required by the data.

\subsubsection{\chandra-only fits}
\label{sec:chandrafits}

To begin, we focused on the \hetg observation.  The source is known to
have significant neutral absorption in the line of sight, with a
column density on the order of $10^{21} \rm cm^{-2}$ (e.g. Dickey \&
Lockman 1990; Kalberla et al 2005).  Therefore,
we first applied an absorbed power law model, utilizing the
\texttt{TBnew} absorption model (an improved version of the
\texttt{TBabs} model of \citealt{wilms:00a}).  This simple model
provides a relatively good fit, yielding a reduced $\chi^2$ value of
1.086 for 3119 degrees of freedom (DoF); however, see
\S\ref{sec:losabs}.  A spectral index of $\Gamma=1.743$ is found.
There are clear residuals in the \fek band near the rest energy of
\feka as expected, as well as in the \meg data below $\sim 1$ keV.
The flux-corrected\footnote{All flux-corrected spectra in this work
  have been created using solely the detector responses, and do not
  reference any assumed models.} \meg and \heg spectra
along with the model fit are shown in Fig.~\ref{fig:initmodelpl}.

Adding a gaussian component to fit the residuals in the \fek band
gives a better fit (reduced $\chi^2=1.063$ for 3116 DoF) and
a similar spectral index ($\Gamma=1.75$).  The rest-frame centroid
of the gaussian was found to be 6.40\,keV, consistent with \feka line emission, 
with a width $\sigma=0.02$\,keV.

Although there appears to be an absorption feature at $\sim7.30$\,keV
in the \heg data (in the rest frame, as seen in the bottom right panel of
Fig.~\ref{fig:initmodelpl}), a significantly improved fit was not
obtained when we attempted to model this feature and we conclude that
it is a statistical fluctuation. We also note that there appears to be
some soft excess below 1 keV, as seen in the \meg spectrum.  We
address this feature in \S\ref{sec:selfcon}.

\subsubsection{Line of Sight Absorption}
\label{sec:losabs}
\begin{figure}
\begin{center}
\includegraphics[width=0.48\textwidth,bb=90 25 597 379]{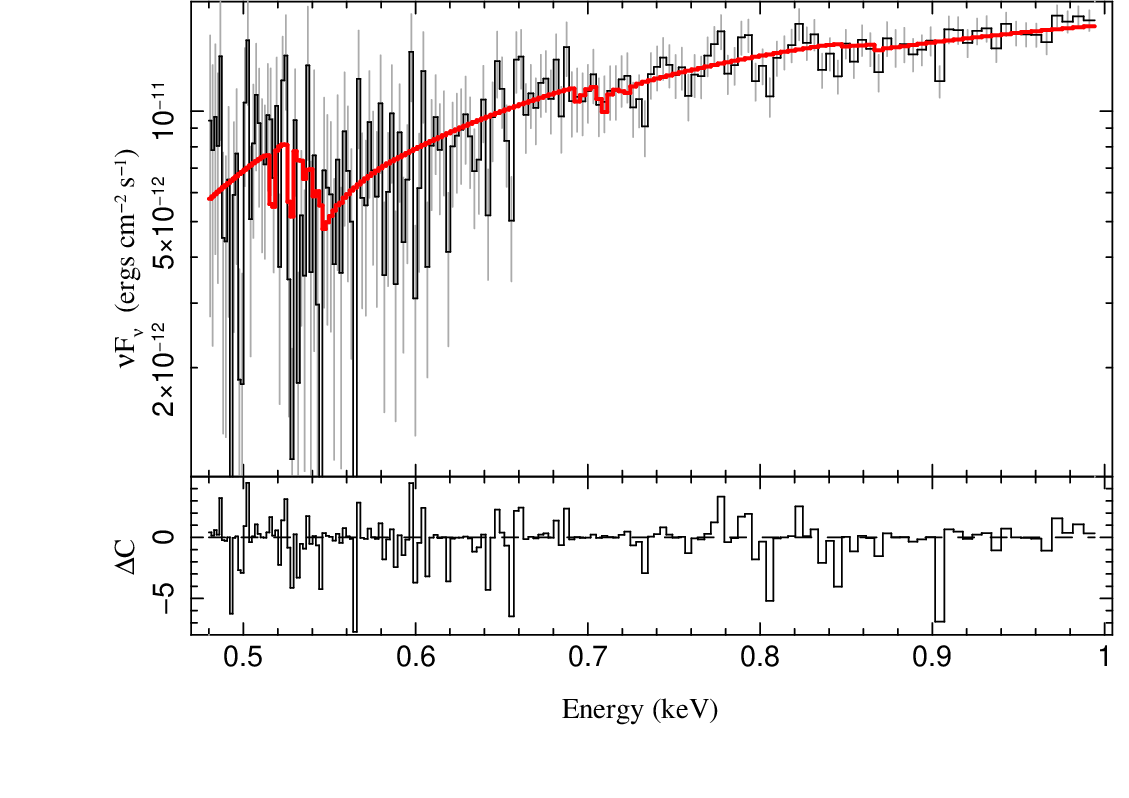}
\end{center}
\caption{A direct fit of the O K-edge, Fe L-edges, and Ne K-edge, as
  observed by the \hetg-\meg in its rest frame, using a simple power
  law and two line-of-sight absorption components both modeled with
  \texttt{TBNew}.  The data are consistent with an upper limit of
  $\frac{2}{3}$ absorption local to \mcgeight.  The summed $N_{\rm H,\ los}$,
  including both extragalactic and Galactic contributions, is
  consistent with that found with Galactic-only absorption.
\label{fig:megedge}}
\end{figure}

We investigated the possibility of neutral absorption in the line of
sight at the redshift of \mcgeight in addition to Galactic
absorption. Considering only the \meg data in the 0.48--1\,keV range
and binning with a uniform 16 bins per channel, we applied two
\texttt{TBnew} absorption models, one with no redshift and the other
placed at the redshift of the source, to a power law continuum and
directly fitted (using Cash statistics; \citealt{cash:79a}) the oxygen
K-edge that is clearly detected in the \meg data.  The \texttt{TBnew}
model also describes the Fe L-edges and Ne K-edge present in the
spectra. The 0.48--1\,keV \meg data and model fit are
shown in Fig.~\ref{fig:megedge}.  Although absorption local to the
system is not required by the data, they are consistent with
$\sim\frac{1}{3}$ (upper limit of $\sim\frac{2}{3}$) of the
line-of-sight absorbing column being located at the redshift of
\mcgeight, with the remaining $\sim \frac{2}{3}$ (lower limit of
$\sim\frac{1}{3}$) being Galactic absorption.  Including two
absorption components gives a slightly better fit: $C = 171$ with 169
data bins, compared to $C =172.9$ with 169 bins for the fit with
Galactic absorption alone.  In this case the total column density of
the two components is consistent with the column density obtained from
fitting with Galactic neutral absorption only (in both cases $N_{\rm
  H,\ los} \sim 0.24\times10^{22}\,\rm cm^{-2}$).  In subsequent fits
described here, for simplicity, we adopted a single neutral Galactic
absorber.

\subsubsection{The Warm Absorber?}
\label{sec:warmabs}
Previous studies of \mcgeight described in \S\ref{sec:intro} find
evidence of warm absorption.  With the exception of the Fe~K region
emission lines described below, we find no evidence of narrow line
structure in either emission or absorption.  To search for narrow line
structure, we employed a Bayesian Blocks technique (based upon the
algorithm of \citealt{scargle:13a} that we have previously used in the
study of \hetg spectra of the low-luminosity AGN, M81$^*$;
\citealt{young:07a}).  Specifically, we fitted a phenomenological
broad-band continuum model to the \meg and \heg spectra (in this case
an absorbed disk+powerlaw+gaussian line), and compared the
\emph{unbinned} model counts and \emph{unbinned} data counts to
determine the optimal binning given a prior significance parameter,
$\alpha$, roughly equivalent to a significance threshold of
$\exp(-\alpha)$.  Significant emission or absorption lines would be
indicated by narrow bins in the optimally partitioned spectrum for
values of $\alpha \aproxgt 3$ (see the description of this procedure
in \citealt{young:07a}). When applied to M81$^*$, this method
identified a significant number of emission lines, and a few
absorption lines, at expected locations, with these line detections
being verified by an independent Monte Carlo detection and fitting
method. For the case of \mcgeight, the Bayesian Blocks search did not
reveal line features in either emission or absorption, in either the
\meg or \heg spectra, at greater than 80\% confidence levels ($\alpha
\aproxgt 1.6$).  This search included  the region surrounding the
"dip" at 7.30~keV, providing further evidence that the feature is not a
real absorption line.  We also performed a similar line search on the
combined \meg and \heg data (unbinned to the \meg resolution, i.e.,
the \heg was binned by a factor of two), and again no line features
were found at greater than the 80\% confidence level.  Our conclusion
is that, in contrast to the prior suggestions, there is no evidence
for a warm absorber in \mcgeight.

\begin{figure}
\begin{center}
\includegraphics[width=0.48\textwidth,bb=90 25 597 379]{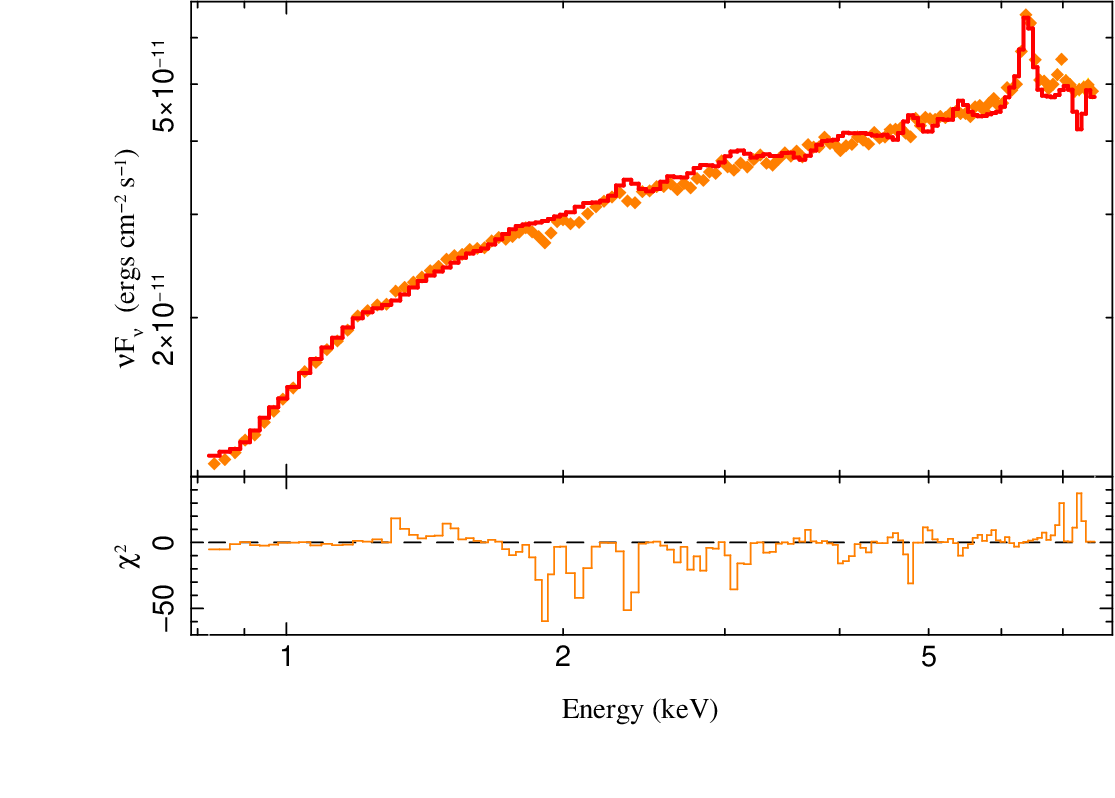}
\end{center}
\caption{A comparison of \chandra-\hetg (red) and \suzaku-\xis (black)
  spectra (see text for details).  Residuals are solely based upon the
  statistical errors of the \suzaku-\xis spectra, and do not include
  the statistical errors of the \heg spectra.
\label{fig:hetgtoxis}}
\end{figure}

\citet{matt:06a} fit a number of absorption lines in the 0.5--1\,keV
range of their \xmm spectra. Although none of these lines are
indicated by the Bayesian Blocks analysis as being significant in the
\hetg spectra, we directly fitted these lines to determine upper
limits to their equivalent widths.  We binned the \meg spectrum by a
uniform factor of 16, and only considered absorbed power law fits over
the 0.48--1\,keV range.  Here we again apply Cash statistics as the
number of counts per bin range only between 0--20 over the
0.48--0.7\,keV band.  The limits on the equivalent width depend upon
the assumed width of the lines (a fitted line width is not given by
\citealt{matt:06a}), and here we fixed the line width to
$\sigma=1$\,eV or $\sigma=5$\,eV.  Fits included only a single
absorption line at a time, with the energy fixed to the best-fit value
reported by \citet{matt:06a}. We took a change of the Cash statistic
of $\Delta C = 2.71$ as a proxy for the 90\% confidence limit for a
single interesting parameter\footnote{For the case of the 999\,eV
  line discussed by \citet{matt:06a} and below, where Gaussian
  statistics applied to the \meg data, there was no difference between
  the confidence limits if one uses a criterion of $\Delta C = 2.71$
  or $\Delta \chi^2 = 2.71$. More properly, one could explore the Cash
  confidence limits with Monte Carlo simulations of the spectra.  We
  have not performed such simulations.}.  With this critereon, the EW
limit is $-2.9$\,eV/$-1.2$\,eV ($\sigma=5$\,eV/1\,eV) for the 526\,eV
line, $-5.4$\,eV/$-1.2$\,eV ($\sigma=5$\,eV/1\,eV) for the 586\,eV
line, $-4.1$\,eV/$-1.3$\,eV for the 627\,eV line
($\sigma=5$\,eV/1\,eV), and $-2.9$\,eV ($\sigma=5$\,eV) for the
654\,eV line.  All of the $\sigma=1$\,eV limits are of smaller
magnitude than found with the fits of \citet{matt:06a}. Only for the
case of $\sigma=1$\,eV for the 654\,eV line was there any improvement
in the fit statistic, where we found an EW of
$-1.6^{+1.4}_{-1.0}$\,eV. This is again of a smaller magnitude than
for the fits by \citet{matt:06a}.  We note, however, that this
marginal detection is consistent with an \emph{unredshifted} {\sc
  Oviii} K$\alpha$ line, as also noted by \citet{matt:06a}, and if
real is unlikely to be associated with the AGN.  We further note that
in the Bayesian Blocks analysis, although not formally significant,
the 654\,eV residual is actually the second most significant narrow
residual, after an only slightly more significant narrow
``absorption'' feature coincident with the unredshifted forbidden line
of Ne{\sc ix}.  This latter feature is not expected in a typical
absorption model.

\begin{figure*}
\begin{center}
\includegraphics[width=0.48\textwidth,bb=90 25 597 379]{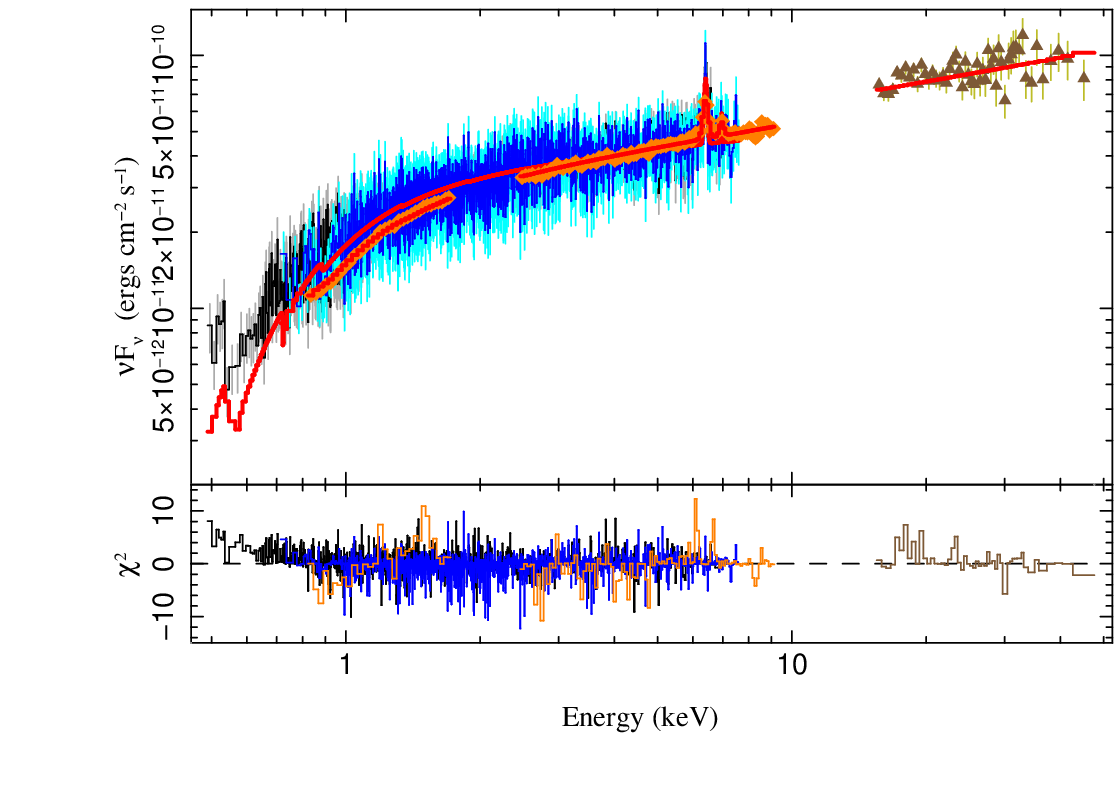}
\includegraphics[width=0.48\textwidth,bb=90 25 597 379]{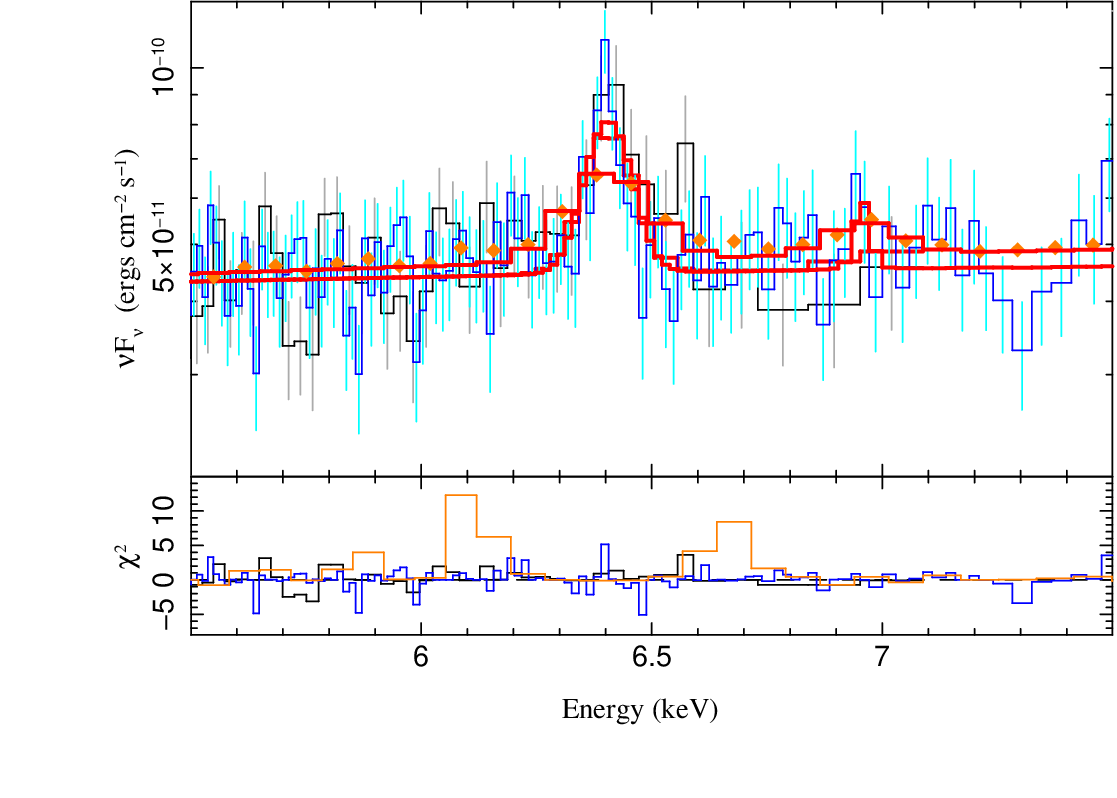}
\includegraphics[width=0.48\textwidth,bb=90 25 597 379]{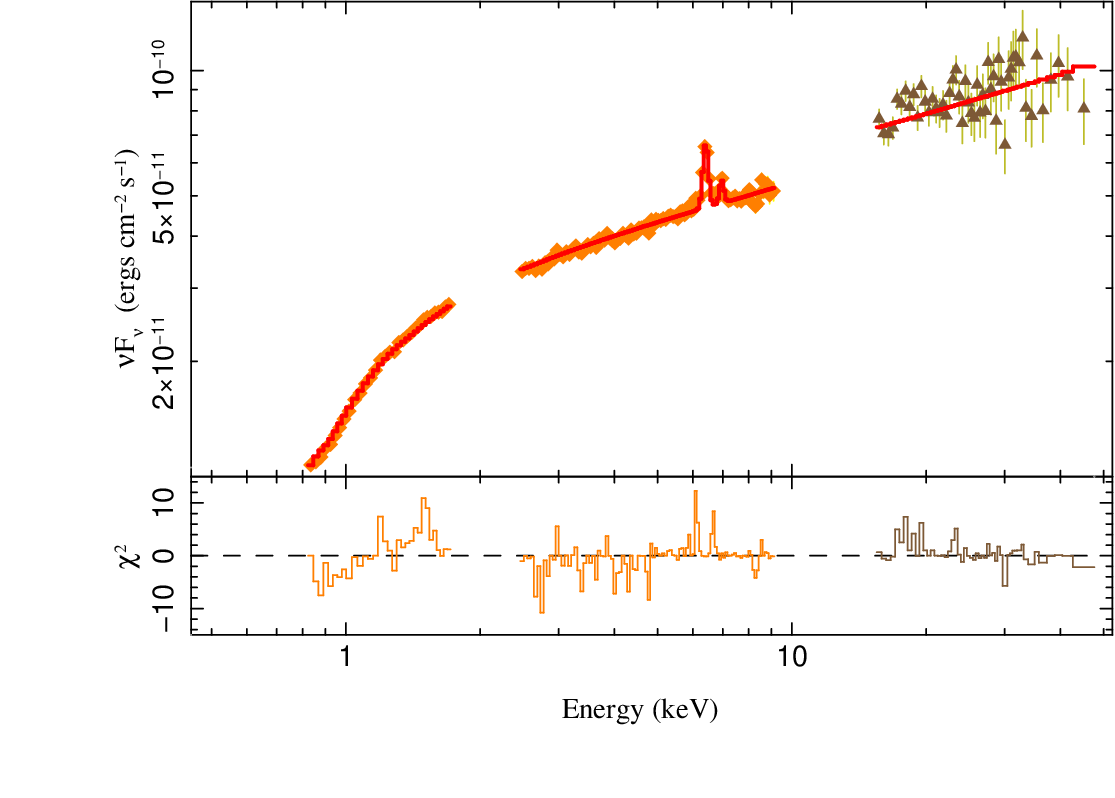}
\includegraphics[width=0.48\textwidth,bb=90 25 597 379]{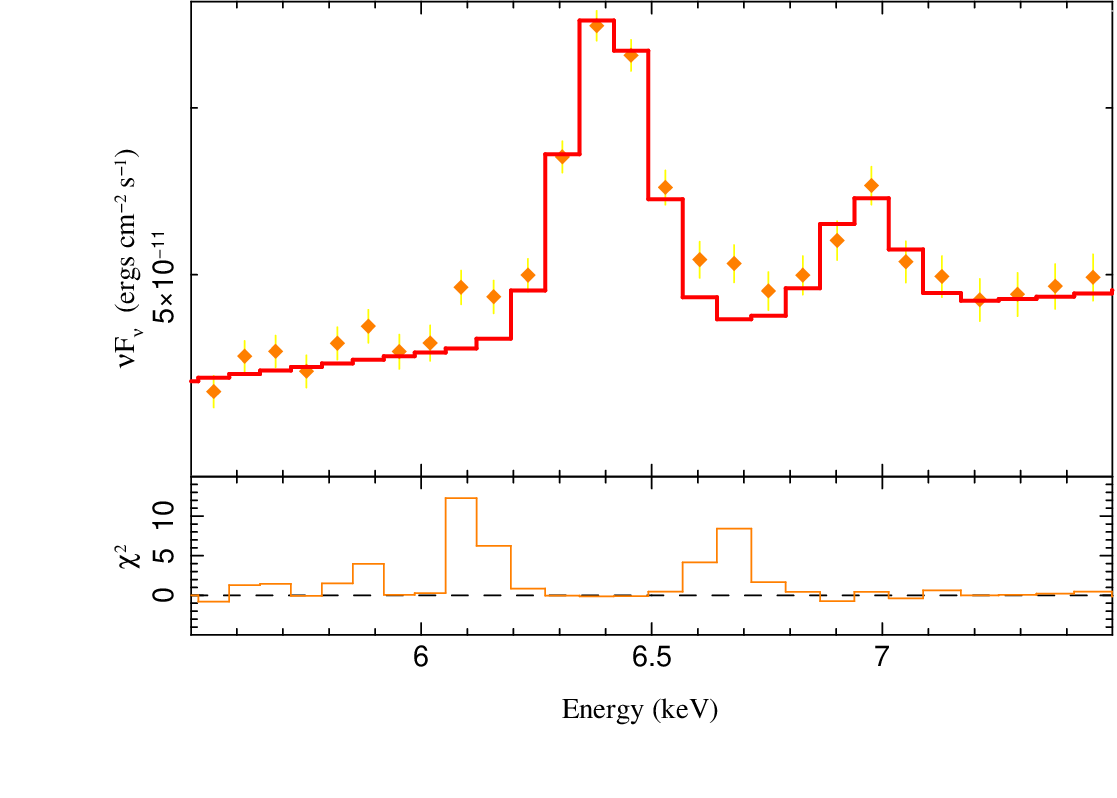}
\end{center}
\caption{Flux-corrected spectra and fit for a phenomenological model
  including an absorbed power-law continuum and \feka and \fexxvi line
  emission.  {\it Top left}: The flux-corrected \meg (black), \heg (blue),
  \xis (orange) and \pin (brown) spectra along with the model fit
  (red).  {\it Top right}: A close-up of the Fe~K region.  {\it Bottom
    left and right}: As above, but showing only the \suzaku spectra to
highlight the differences in the high-resolution gratings spectra and
the CCD spectra.  The right panels show residuals in the CCD 
spectrum that could indicated an \feka Compton shoulder and/or
additional broad line emission.  The left panels show soft excess
below $\sim 1$  keV that is only probed by the \meg data.
\label{fig:initmodelall}}
\end{figure*}

For the 999\,eV line discussed by \citet{matt:06a}, we can also use
the \heg data.  Given the larger number of counts near this energy
region, we applied a uniform binning of a factor of 4 to the \meg
data, and a uniform binning of a factor of 8 to the \heg data, and
fitted an absorbed power law (again using Cash statistics) in the
0.8--1.1\,keV band.  The EW limit is $-1.3$\,eV/$-0.2$\,eV
($\sigma=5$\,eV/1\,eV) for an absorption feature at this location.
Again, this is of smaller magnitude than the absorption line fits of
\citet{matt:06a}.  We conclude that there is no compelling evidence
for a warm absorber in the \hetg spectra of \mcgeight.

\subsection{\chandra + \suzaku fits}
\label{sec:chandsuzprelim}
It was necessary to expand our data set to include the \suzaku-\xis
and \pin spectra in order to further constrain the continuum and the
emission in \fek band.

The \chandra-\heg and \meg and the \suzaku-\xis and \pin spectra both
reveal a typical absorbed power-law shape continuum and strong, narrow
emission at the rest energy of \feka.  In Fig.~\ref{fig:hetgtoxis} we
show a comparison of the \hetg-\heg data to the \xis data, where we 
flux-corrected the combined \heg spectra and
folded them through the \suzaku response, allowing for an overall
``tilt'' in spectral slope (centered at 3\,keV) and cross-normalization
constant.  We found that, although the \chandra and \suzaku data are
not simultaneous, their similar shape facilitates global
fitting. Specifically, the change in spectral slope was $\Delta \Gamma
\approx 0.046$ with an \heg to \xis cross normalization constant of
$\approx 0.96$.  These are within the likely cross-calibration
differences of the two instruments \citep{ishida:11a}. (The residuals
in the \xis spectra are predominantly known, systematic effects in the
\suzaku responses.)  Given the similarity of these spectra, we
proceeded to include all of the \chandra-\hetg and \suzaku-\xis and
\pin data in subsequent broadband fits, allowing a constant
normalization factor between the fits to the data from the two
missions.  Note that the dip feature near 7.30\,keV in the \heg
spectrum is not seen in the \xis spectrum; however, we have not
accounted for the statistical uncertainties in the \heg data in this
comparison.

We applied the phenomenological model consisting of a power law
spectrum, Galactic absorption, and \feka line emission (as described
above) to the four sets of spectra.  This model yielded a worse fit
than that found by fitting the \chandra data alone, giving a reduced
$\chi^2$ value of 1.140 for 3297 DoF.

\begin{figure}
\begin{center}
\includegraphics[width=0.48\textwidth,bb=90 25 597 379]{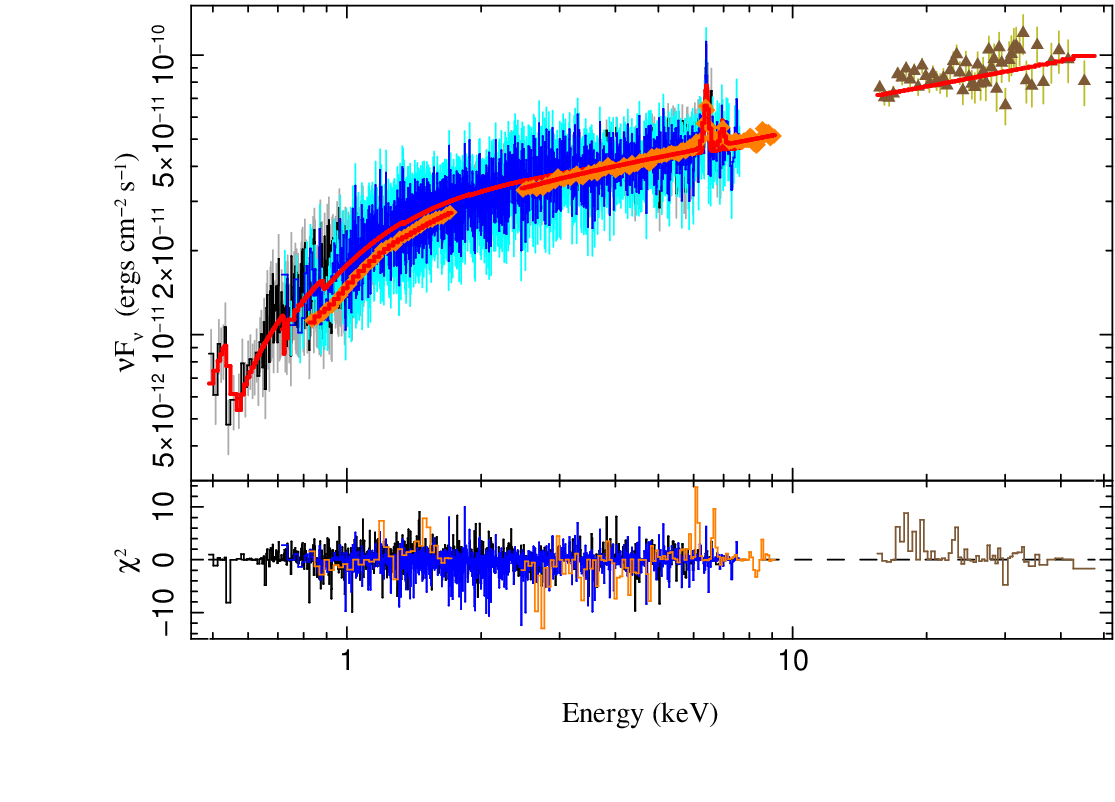}
\end{center}
\caption{As described in Fig.~\ref{fig:initmodelall}, with a soft
  excess component, phenomenologically described with \texttt{diskbb},
  added to the model.
\label{fig:initmodeldbb}}
\end{figure}

Residuals consistent with \fexxvi line emission are present in the
\xis spectra (as expected from the results of previous studies; see
\S\ref{sec:intro}) that are not evident in the \hetg data.  When we
attempted to add a gaussian line component to model \fexxvi line
emission in the \hetg data alone, the centroid energy settled at 6.31
keV in the rest frame, with a line width $\sigma=126$\,eV. The added
component may in fact have been attempting to model either a Compton
shoulder for the \feka emission line or underlying broadened \feka
line emission.  We consider both of these possibilities in
\S\ref{sec:selfcon}.  Adding this gaussian component to fit all four
sets of data simultaneously (see Fig.~\ref{fig:initmodelall}), on the
other hand, resulted in a slightly better fit (reduced $\chi^2=1.123$
for 3293 DoF), and the best-fit centroid energy of the gaussian was
consistent with \fexxvi line emission.  The top, left panel of Fig.~
\ref{fig:initmodelall} shows the flux-corrected \meg, \heg,
\xis, and \pin spectra along with the model fit.
A close-up of the \fek band is shown on the right.  For clarity we
also show only the flux-corrected \suzaku spectra and model in the bottom
panels.  The \suzaku-\xis spectrum shows residuals redward of the
\feka line emission that could indicate a Compton shoulder and/or
additional broad line emission.  The \chandra-\hetg spectrum shows
additional soft excess in the continuum below $\sim 1$\,keV.  Adding a
thermal disk component (\texttt{diskbb}) to phenomenologically model
the soft excess improves the fit (reduced $\chi^2=1.080$ for 3290
DoF), as seen in Fig.~\ref{fig:initmodeldbb}.  Although there are
residuals in the \xis and \meg spectra that appear to be consistent
with \fexxv line emission, we did not find that an added gaussian was
required by the data.

As expected, utilizing both the \chandra and \suzaku data yields a
more robust picture of the detailed emission and absorption mechanisms
of the source since each of the instruments is suited to probe
particular features in the spectrum.  This preliminary investigation
showed a combination of several possible spectral components,
including an intrinsic power-law shaped continuum, neutral
line-of-sight absorption, soft excess, \fexxvi line emission, and
Compton reflection (including \feka line emission with a possible
Compton shoulder and/or broad \feka line emission plus possible
continuum curvature).  We explore these features with more
self-consistent models in the following section.

\subsection {Modeling with Self-Consistent Components}
\label{sec:selfcon}
The combination of high spectral resolution \chandra-\hetg data with
broadband, high signal-to-noise \suzaku data allows us to constrain
the physical parameters of the system by applying self-consistent
X-ray spectral fitting models that have been recently developed.
Physically-based spectral fitting models offer a clear advantage over
purely phenomenological ones as they allow us to extract more robust
and physically-meaningful constraints on each of the regions of an
AGN, as well as the system as a whole.  Our best-fit model consists of
a power-law intrinsic continuum, soft excess (\texttt{diskbb},
included as a purely phenomenological component), \fexxvi line
emission, Galactic absorption (\texttt{TBNew}), and reprocessing by
circumnuclear material.  Below we describe the results of two sets of
fits that utilize different models for the last component.  In all of
the fits, we allowed for some spectral variability between the
\chandra and \suzaku observations.  In particular, we did not tie the
intrinsic power law continua, nor the normalizations of the soft
excess and the \fexxvi line emission in the joint fits of those data
sets.

Employing sophisticated models such as those discussed below is
computationally expensive, especially when most of the model
parameters are allowed to remain free.  We therefore found it
expedient to utilize a Monte Carlo Markov Chain (MCMC) code to fit the
models to the data and to explore the parameter space as rigorously as
possible.  The code was developed specifically for ISIS, based upon
the parallel ``simple stretch" method from \citet{foreman:13a} which
in turn was based on the work of \citet{goodman:10a}.

For each MCMC run we created 200 (400) ``walkers'' which were evolved
for 4700 (2500) steps for a total of $7.4 \times 10^5$ ($3 \times
10^6$) samples for the \myt (\pexmon) model.  Histograms and error
bars are based on the last half of these samples (i.e., the first half
were discarded as part of the MCMC ``burn in''.)

\subsubsection{Modeling with \pexmon}
\label{sec:pex}
It is clear from our phenomenological fitting and from the literature
(\citealt{matt:06a}, \citealt{bianchi:10a}, \citealt{soldi:10a}) that
\mcgeight likely includes a reflection component due to distant, cold,
circumnuclear material.  For comparison with previous results, in
particular from \citet{bianchi:10a}, and the results of fits with the
\myt model (see \S\ref{sec:myt}), we modeled the reprocessor with
\pexmon \citep{nandra:07}.  \pexmon is a
self-consistent model that includes Compton reflection and fluorescent
line emission (\feka, \fekb, and \nika)
from a slab of neutral material subtending a solid angle of $2\pi$ at the X-ray source (to represent
reflection from an accretion disk, although it is sometimes used to describe 
reflection from a region such as the torus or BLR, as we do here for
comparison with previous work).

\begin{figure*}
\begin{center}
\includegraphics[width=0.48\textwidth,bb=90 25 597 379]{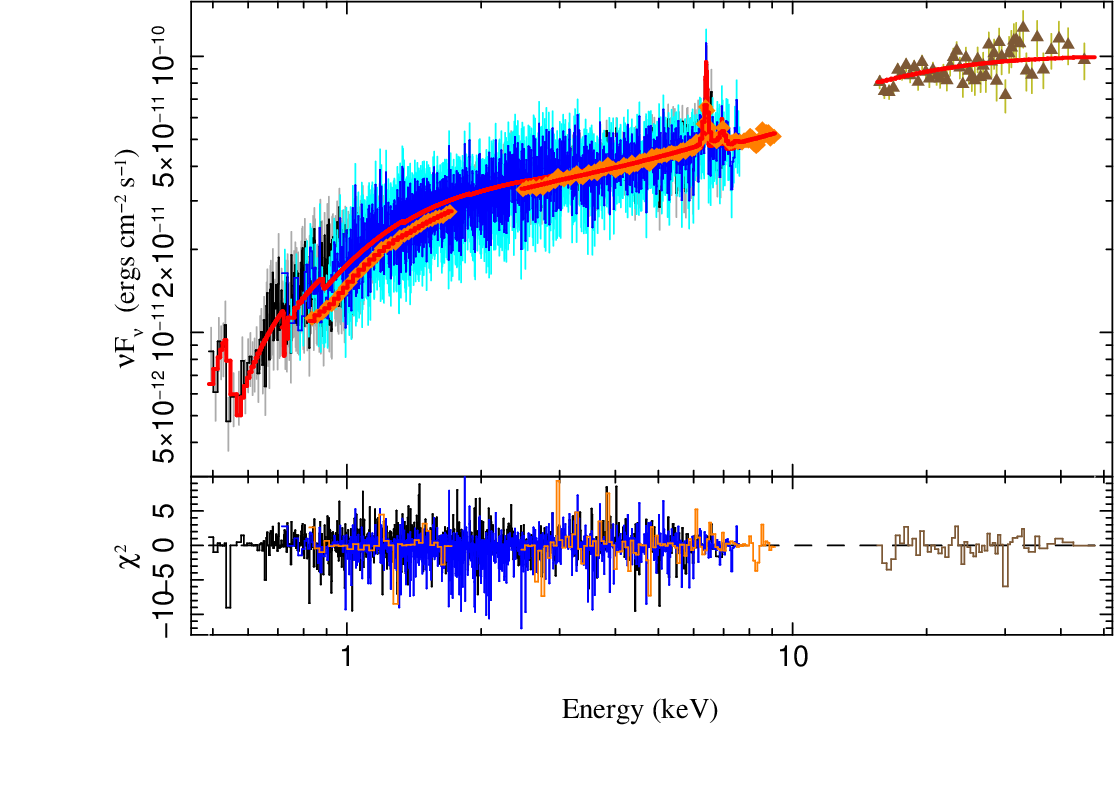}
\includegraphics[width=0.48\textwidth,bb=90 25 597 379]{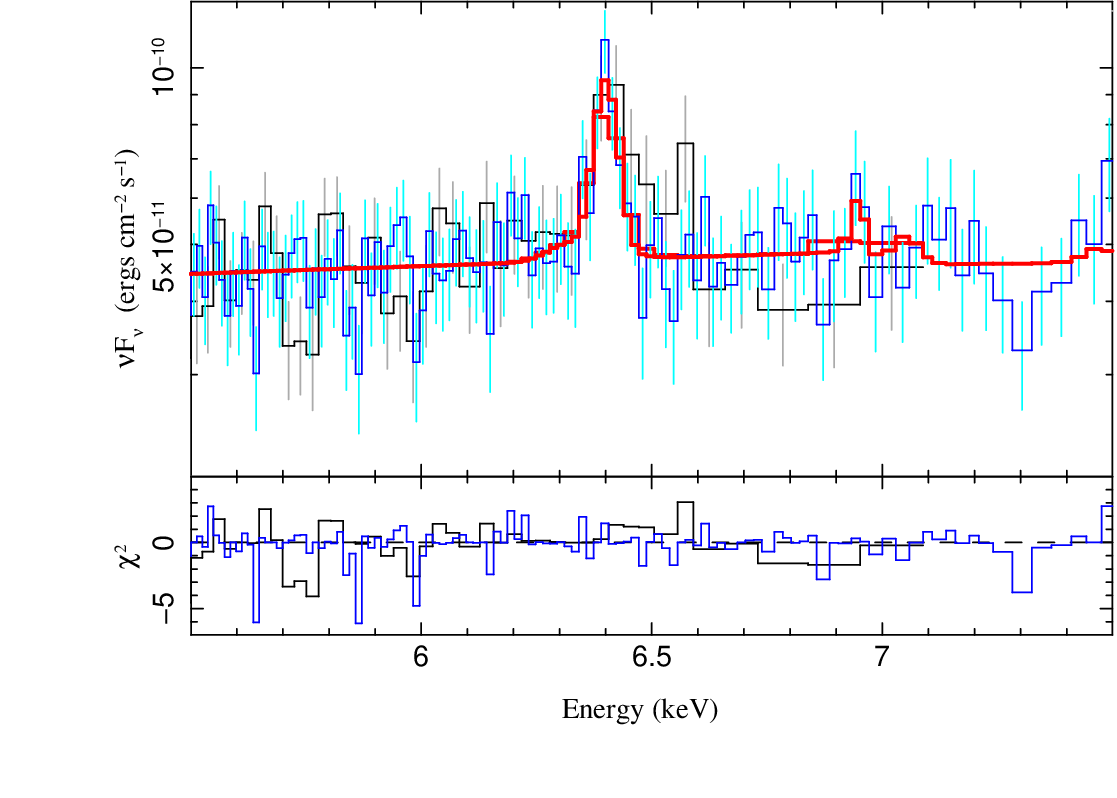}
\end{center}
\caption{Flux-corrected spectra and fit for a model including an absorbed
  power-law continuum, soft excess, \fexxvi line emission, and cold
  Compton reflection (\pexmon).  {\it Left:} The flux-corrected \meg
  (black), \heg (blue), \xis (orange) and \pin (brown) spectra along
  with the model fit (red).  {\it Right:} A close-up of the Fe~K
  region for the \hetg data only.
\label{fig:pexfit}}
\end{figure*}

The best-fit model spectrum is shown in Fig.~\ref{fig:pexfit}.  The
MCMC model parameters (reduced $\chi^2=1.048$ for 3288 DoF) and 90\%
confidence errors are given in Table~\ref{table:mytpexresults}.  We
obtained intrinsic power-law spectral indices of
$\Gamma=1.897^{+0.025}_{-0.025}$ (\chandra) and
$\Gamma=1.818^{+0.025}_{-0.015}$ (\suzaku).  Although not tightly
constrained, the 90\% confidence range for the high-energy cutoff
($E_{\rm C}=155^{+223}_{-106}$) was consistent with the reported
values measured by \swift-BAT, \bsax, {\it OSSE}, and \integral.  The
neutral line-of-sight column density was slightly higher than that
found in our preliminary analysis (\nh$\sim0.39\times10^{22}$ as
opposed to $0.24\times10^{22}\,\rm cm^{-2}$), see \S\ref{sec:losabs}.
We modeled the soft excess, which was clear in the \meg residuals
below $\sim 1$\,keV, with a simple disk component (\texttt{Diskbb}),
finding a peak temperature of $kT=0.091^{+0.007}_{-0.006}$\,keV.

\begin{figure*}
\begin{center}
\includegraphics[width=0.33\textwidth,bb=90 25 597 379]{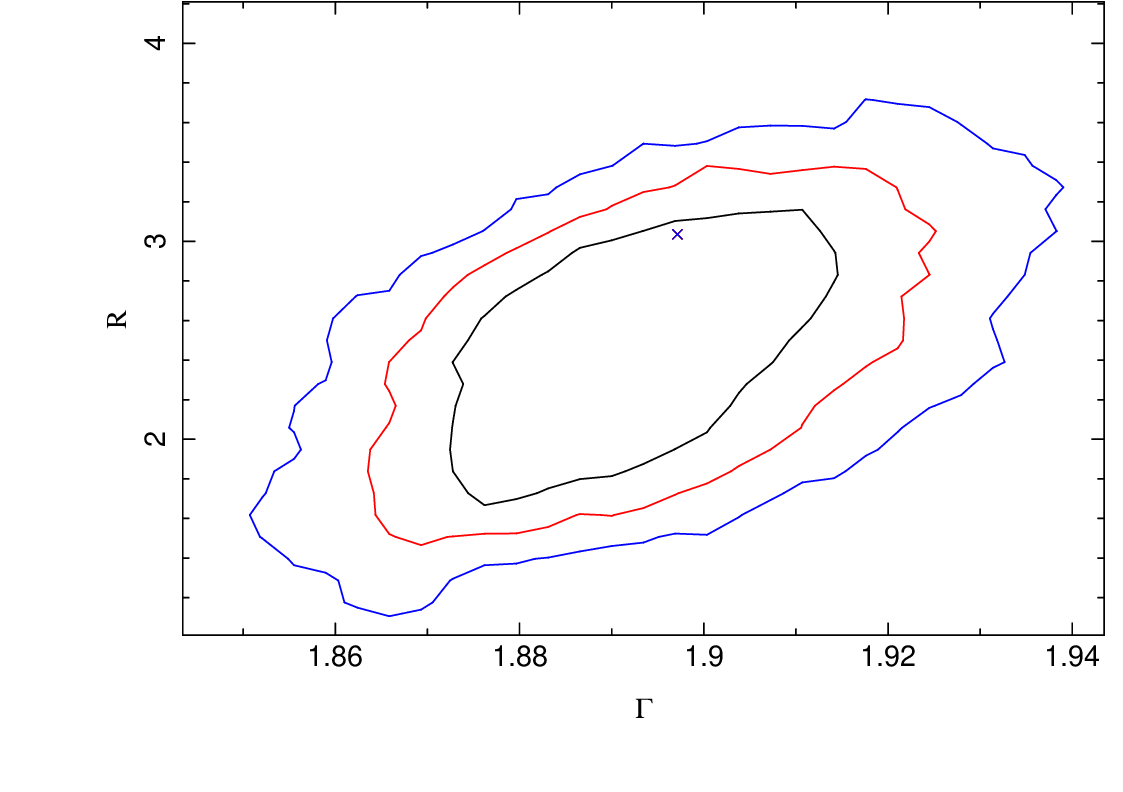}
\includegraphics[width=0.33\textwidth,bb=90 25 597 379]{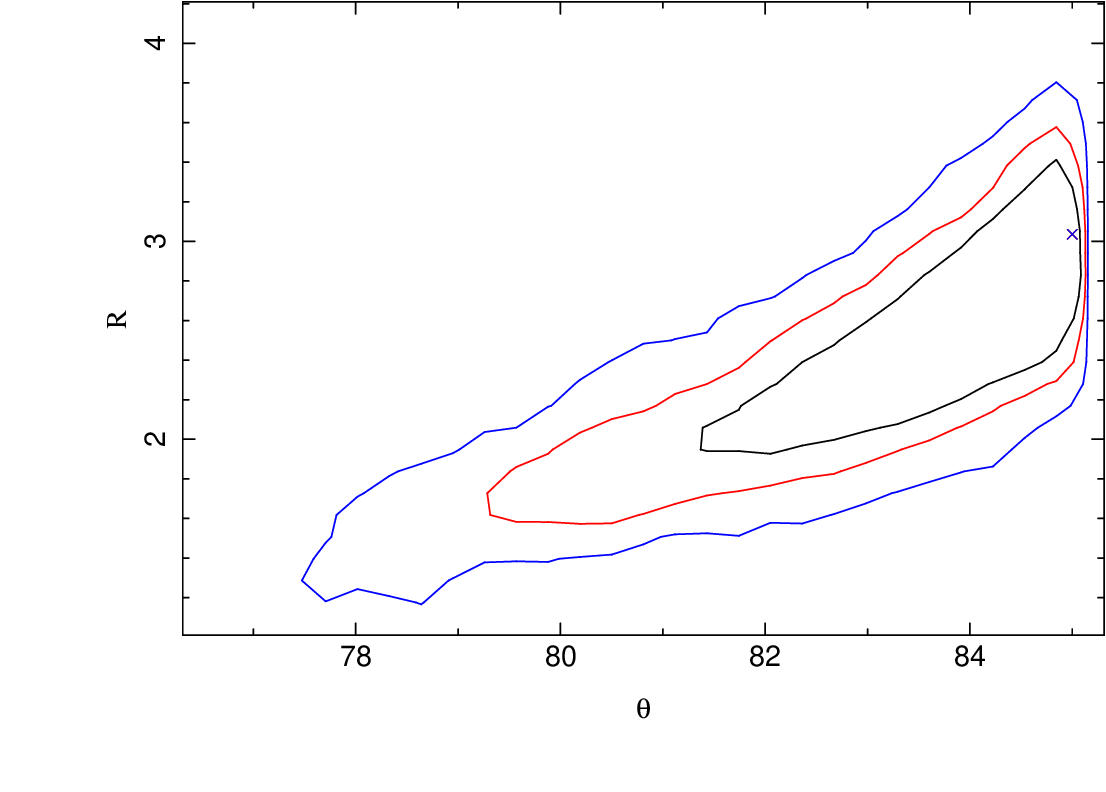}
\includegraphics[width=0.33\textwidth,bb=90 25 597 379]{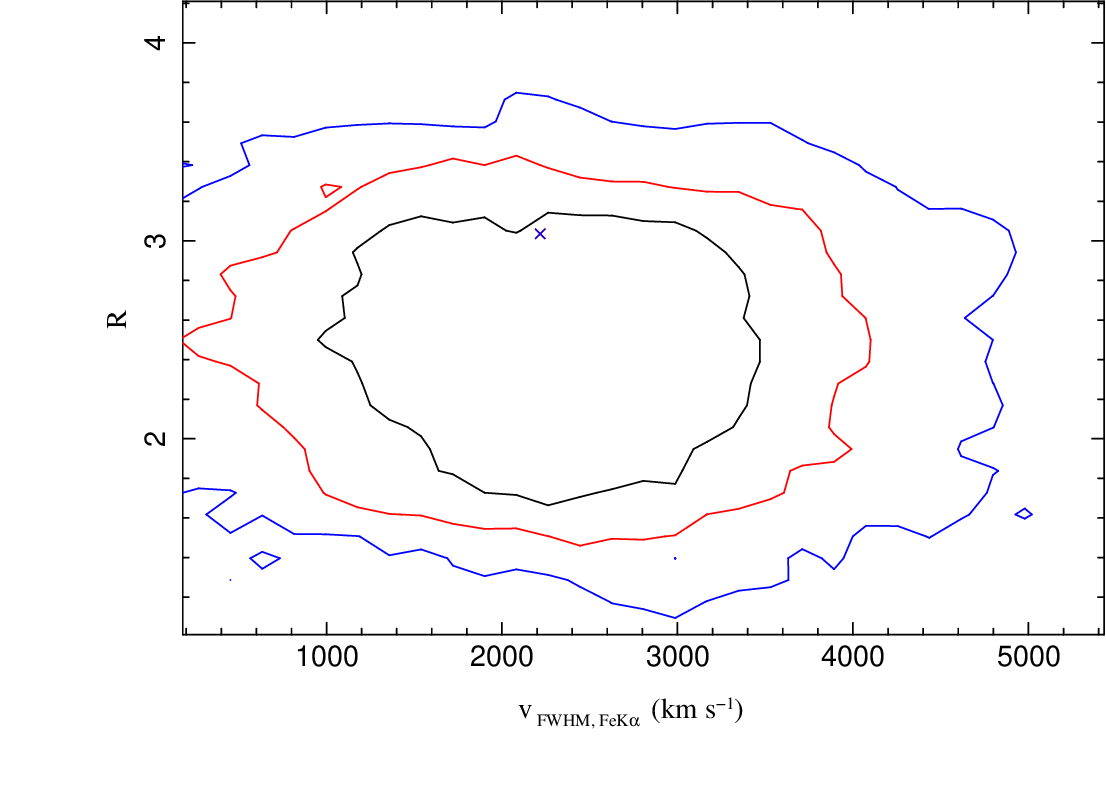}
\end{center}
\caption{The 68\% (black), 90\% (red) and 99\% (blue) 
confidence contours for the model including \pexmon.
  Shown are contours for the reflection fraction versus the spectral
  index of the \hetg data ({\it left}), the inclination
  angle of the reflector ({\it center}), and the FWHM velocity width of the
  \feka emission line.  
\label{fig:pexcont}}
\end{figure*}

The Compton reflection fraction, $R = 3.04^{+0.42}_{-1.26}$, was
higher than that reported by \citet{bianchi:10a} from their
analysis of the \suzaku-\xis+\pin data.  The inclination angle was found to be high,
pegging the model limit of $\theta=85$.  The 68, 90, and 99\%
confidence contours of the reflection fraction versus photon index
({\it left}) and inclination angle ({\it center}) in
Fig.~\ref{fig:pexcont} shows the degeneracy of these parameters.  In
general, we see that lower inclination angles also allow for lower
reflection fractions.  It is important to point out, however, that the
physical meaning of $R$, when this model is used in this capacity,
is not clear.  As discussed in MY09, there is no correlation between $R$ 
and the column density of the distant reflector.  $R$ cannot simply 
be interpreted as a fraction of the subtended solid angle $2\pi$, nor 
can it give a clear indication of reflection "strength" since that is geometry
dependent.

To 90\% confidence, there is a large range in the abundance of Fe,
with $A_{\rm Fe}=0.65$--1.09.  The \pexmon model was convolved
with \texttt{gsmooth} to include the affects of kinematics.  From this
we found the width of the \feka emission line at 6.4\,keV to be
$\sigma=19^{+11}_{-13}$\,eV, with a corresponding range in full width
at half maximum velocity of $\sim950$--7000\,km/s, assuming Doppler
broadening.  This implies a distance of 0.014--0.76\,pc to the Fe~K
line emitting region from the black hole, assuming Keplerian motion
and a black-hole mass of $M_{\rm BH}=1.2\times10^{8}M_{\odot}$
\citep{winter:10a}.  In Fig.~\ref{fig:pexcont} ({\it right}) we show
confidence contours of the reflection fraction versus FWHM velocity
width of the \feka emission line.

\begin{figure}
\begin{center}
\includegraphics[width=0.48\textwidth,bb=90 25 597 379]{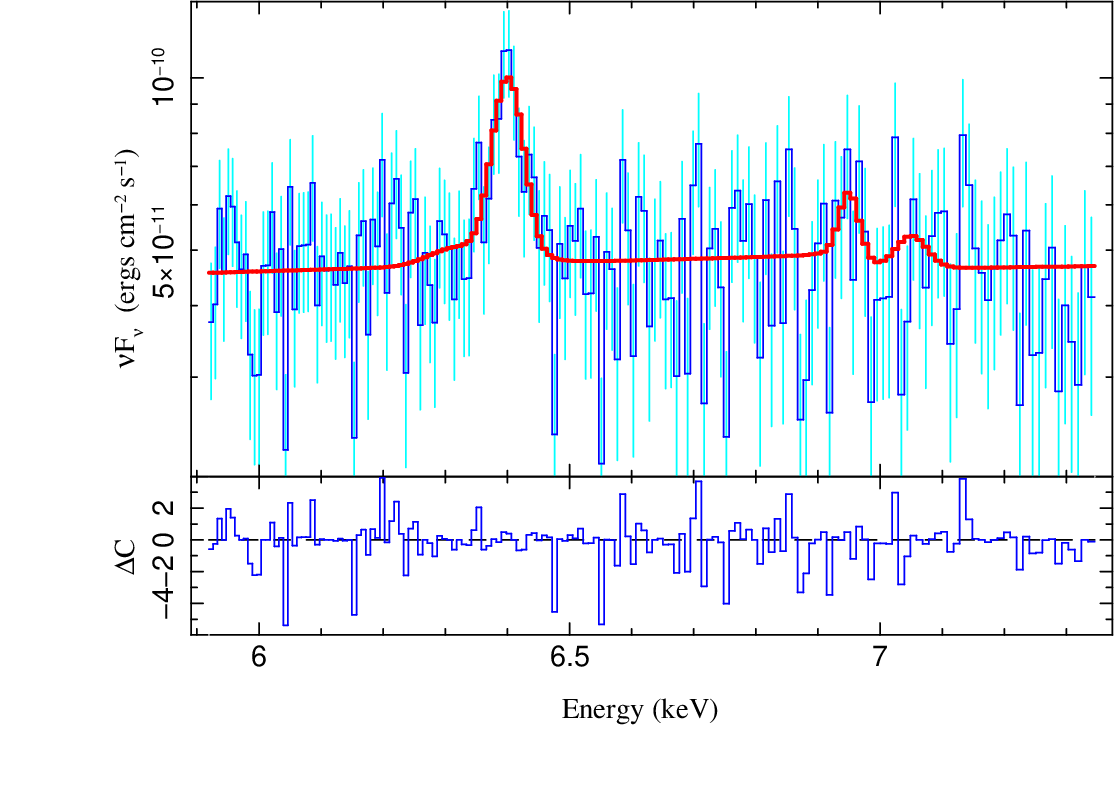}
\end{center}
\caption{The flux-corrected \heg spectrum (blue) and fit (red) to the Fe~K
  band for the model including \pexmon.
\label{fig:hegpex}}
\end{figure}

For comparison, we considered only the \chandra-\heg data in the Fe~K
band using the same model.  We further restricted the energy range to
5.8--7.2\,keV, unbinned the data to achieve the maximum resolution of
the \heg, and fit the model using Cash statistics.
Figure~\ref{fig:hegpex} shows the \heg data and the model fit in this
band.  In this case, the \feka emission line is resolved, with a
best-fit value of the \feka emission line is $\sigma =
17^{+11}_{-9}$\,eV, consistent with the global fit.

\subsection {Relativistic Disk Emission?}

Although the Fe~K line emission and associated Compton hump appear to
be well described by neutral reflection that is not relativistically
smeared, there may be degeneracy between this and a relativistic
component.  We therefore included relativistically smeared, ionized
reflection from an accretion disk ({\tt reflionx}, \citealt{ross:05a},
convolved with {\tt relconv}, \citealt{dauser:10a}) to investigate the
contribution to the reflection continuum and Fe~K line emission.
Since broad Fe~K line emission is better constrained by \suzaku, as a
first step we considered those data alone.  We tied the inclination
angle of the \texttt{relconv} smearing to that of the \pexmon
component; however, we fixed the inner radius of the smearing to that
of the marginally stable orbit for the fitted black hole spin ($a^*$)
and the outer smearing radius to $400\,r_{g}$. We fixed the cutoff
energy of the power-law in the \pexmon model to be $300$\,keV (i.e.,
the value fixed in the \texttt{reflionx} model). We further tied the
\pexmon and \texttt{reflionx} power-law indices and Fe abundances
together, but left the \texttt{reflionx} ionization parameter to be
freely fit.

We obtained a good fit to the data ($\chi^2=1.048$ for 169 DoF), with
similar results to the \chandra-\hetg+\suzaku fits without a
relativistically smeared reflection component.  The inclination was
again found to be pegged at the \pexmon limit of $85^\circ$.  We found
a slightly harder power-law slope ($\Gamma=1.78^{+0.05}_{-0.03}$) and
a similar reflection fraction ($R\sim2.0^{+0.8}_{-0.5}$) compared to
the fit excluding relativistically smeared reflection from the
accretion disk (see Table~\ref{table:mytpexresults}).  The major
difference for this model is that the fit implied an overabundance of
Fe ($A_{Fe}=1.8^{+0.9}_{-0.8}$).

\begin{figure*}
\begin{center}
\includegraphics[width=0.48\textwidth]{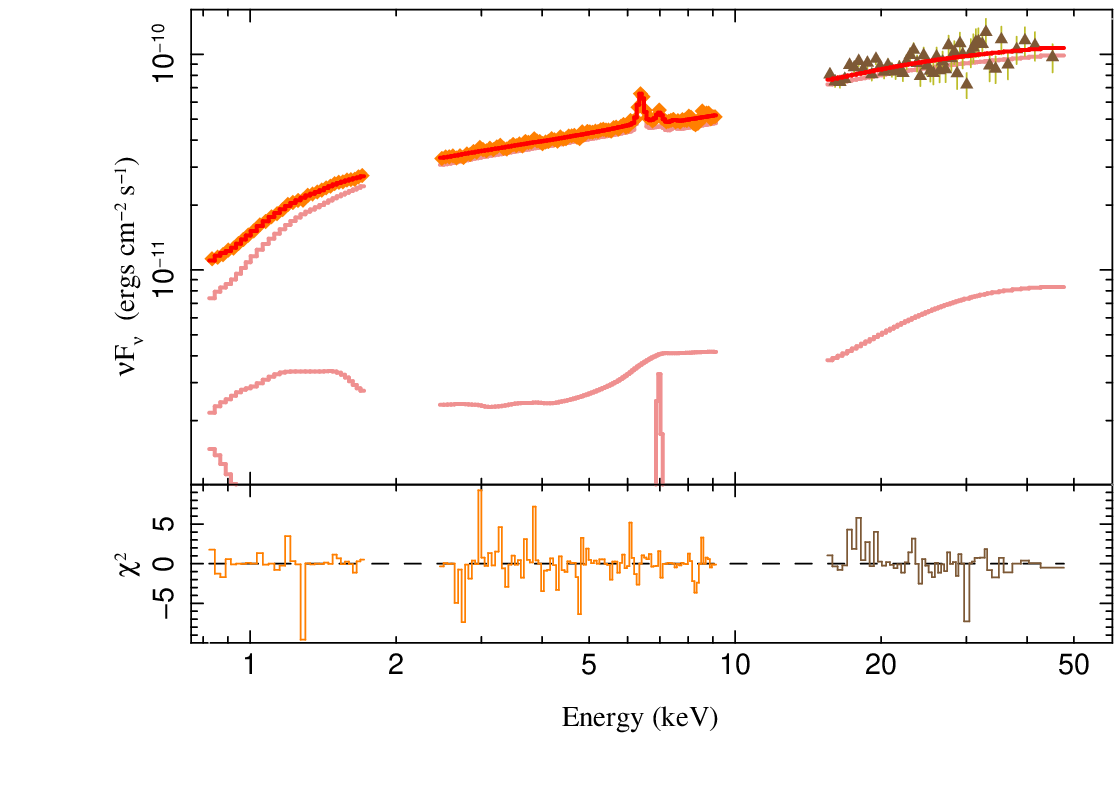}
\includegraphics[width=0.48\textwidth]{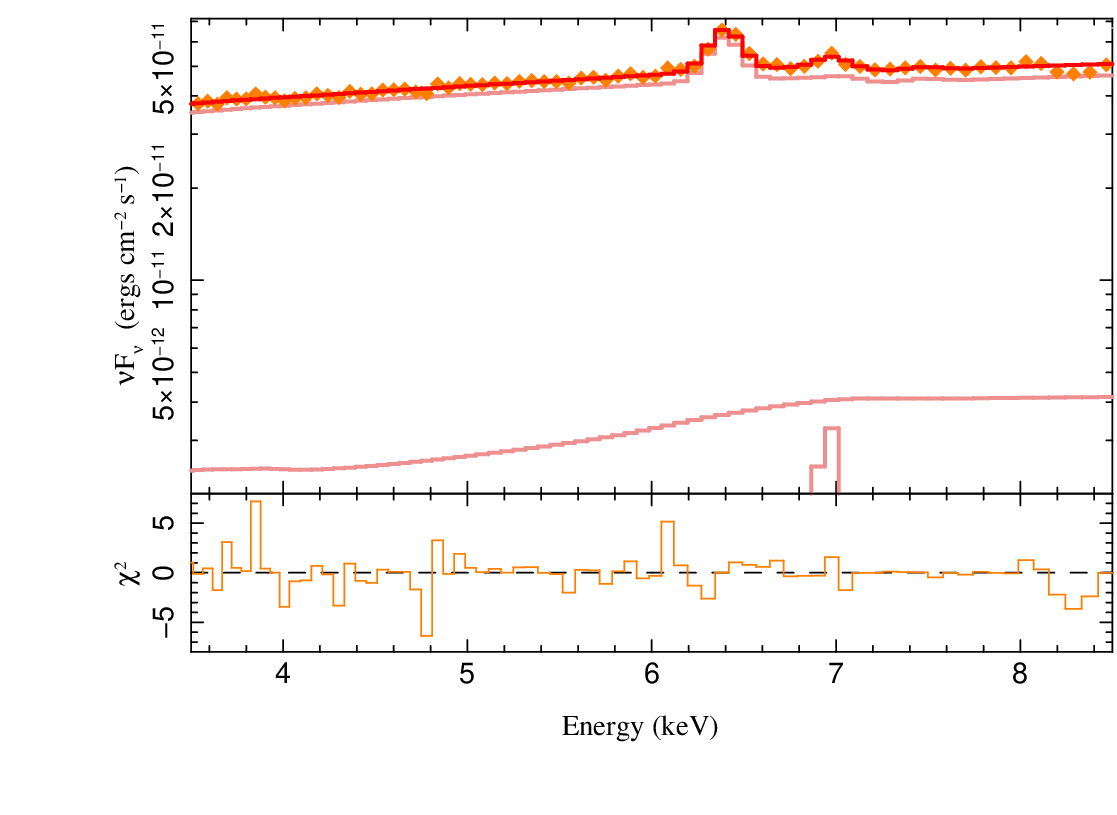}
\end{center}
\caption{{Left}: \suzaku only fits with relativistically smeared
  ionized reflection. The individual \pexmon, soft excess,
  relativistically smeared reflection, and \fexxvi components are
  shown.  {Right:} Close up of the Fe~K band, also showing the
  individual model components.
\label{fig:suzpexrel}}
\end{figure*}

As regards the relativistic smearing parameters, the best fit spin was
found to be at the maximal prograde limit of the \texttt{relconv}
model, $a^*=0.998$, with the lower-limit being $a^* \ge 0.799$.  The
emissivity index for the relativistic smearing is $2.1^{+4.9}_{-0.3}$,
i.e., consistent with being (but not required to be) dominated by
emission from near the marginally stable orbit. For the lower range of
the emissivity index, the width of the contributed \feka line is in
fact dominated by the ionization (with the fitted ionization parameter
being $\Xi = 570^{+480}_{-240}$), rather than the relativistic smearing.
The overall contribution of this reflection component is small, as
seen in Fig.~\ref{fig:suzpexrel}.  Over the $\approx 4$--9\,keV band
it contributes $\approx 5\%$--8\% of the flux, and allows for a small
degree of extra curvature in the Fe line region.  The overall
normalization of the \texttt{reflionx} component is formally greater
than zero at the 90\% confidence level; however, we note that removing
this component completely (i.e., removing four fit parameters: ionized
reflection normalization and ionization parameter, black hole spin,
and relativistic smearing emissivity index) only alters the best fit,
absolute $\chi^2$ by 8.5, with the fitted Fe abundance again becoming
lower ($A_{Fe}=0.9\pm0.2$).  As pointed out by \cite{matt:06a}, the
existence of this component primarily serves to produce a reflection
hump, without producing an associated narrow \feka line component,
thus allowing for higher Fe abundances in the unsmeared reflector.

Although the relativistically smeared, ionized reflection model does
fill in some of the curvature around the \feka emisison line in the
\suzaku data, its presence is not strongly required by the data.  We
therefore did not include relativistic disk emission in subsequent
fits.

\subsubsection{Modeling with \texttt{MYTorus}}
\label{sec:myt}
We completed a similar analysis to that described in \S\ref{sec:pex},
using \texttt{MYTorus} (\citealt{murphy:09b}\footnote{See
  www.mytorus.com for more details.}; hereafter MY09) instead of
\pexmon to model the cold reprocessor. \texttt{MYTorus} is a fully
relativistic 3D model that assumes a toroidal shape, but may also be
used to model a cloud-like or ``clumpy" geometry (see
\citealt{yaqoob:12a} for details).  The model does not presuppose a
fixed distance from the nucleus nor does it assume the material is
Compton thick and therefore, for example, could be used to approximate
absorption and reflection from the BLR.  As \myt self-consistently
models absorption, reflection, and \feka and \fekb line emission
(including the associated Compton shoulders), it is more suitable for
constraining the observed reflection component than purely
phenomenological modeling.  For this fit, we included \nika line emission
with a separate Gaussian component.  Since the time-averaged signatures of the
torus component are assumed to remain roughly constant, we do not
allow for variability of this component between the \chandra and
\suzaku observations.  For the following, we set the relative
normalization parameters of the scattered and line spectra to unity.

We assumed a terminal energy of 500\,keV.  The terminal energy of the
\myt model is not the same as an observed cutoff in the power law
continuum but refers to the highest energy assumed for the intrinsic
photons that were incident to the torus prior to
absorption/reflection.  The observed cutoff, which has previously been
reported to be $101^{+48}_{-20}$\,keV (\integral and \swift
\citealt{soldi:10a}), $150^{+30}_{-20}$\,keV (\swift;
\citealt{bianchi:10a}), $170^{+300}_{-80}$\,keV (\bsax;
\citealt{perola:00a}), and $270^{+90}_{-70}$\,keV (\osse;
\citealt{grandi:98a}), can in fact be produced by Compton scattering
of higher-energy photons in a Compton-thick medium (see MY09).

For comparison we fitted the data utilizing a 200\,keV terminal energy
and found the same reduced $\chi^2$ and consistent model parameters.
Although we did not expect this difference in terminal energy to
significantly affect the band that we fitted, it would
potentially affect fits to higher energy spectra (e.g., \swift).  We
defer an analysis including such data to later work.  In the following
we report results from fits utilizing the 500\,keV terminal energy
\myt model tables.

Considering the degeneracies in the model parameters (e.g., between
the photon index [$\Gamma$], the equatorial column density [\nh] of
the torus, and the inclination angle [$\theta$] of the torus), it is
necessary to first coarsely and methodically explore the parameter
space in order to avoid settling into a local minimum in $\chi^2$.  We
thus determined that a face-on inclination angle (i.e., the torus is
seen in reflection only) near $\theta\sim40^{\circ}$ and an equatorial
column density of approximately \nh$\sim10^{24} \rm cm^{-2}$ is
preferred by the model\footnote{For example, for fits where $\theta$
  was fixed to $15^{\circ}, \ 30^{\circ}, \ 45^{\circ}, \ 65^{\circ},
  \ {\rm and} \ 85^{\circ}$, \nh was $\rm 1.42, 1.05, 1.16, 0.01, and
  \ 0.01 \times 10^{24} \ cm^{-2}$ and the reduced $\chi^2$ was 1.380,
  1.057, 1.385, 1.259, and \ 11.389 respectively.  Both edge-on cases
  were pegged at the lower limit of the model in \nh.  In the face-on
  cases, $\Gamma$ remained in the approximate range 1.75--1.85.  For
  $\theta=65^{\circ}$ and $85^{\circ}$, $\Gamma=1.69$ (1.66) and 1.61
  (1.44) for the \chandra (\suzaku) data and the fit to the \fek band
  was (visibly) obviously inaccurate.}.

\begin{figure*}
\begin{center}
\includegraphics[width=0.33\textwidth,bb=90 25 597 379]{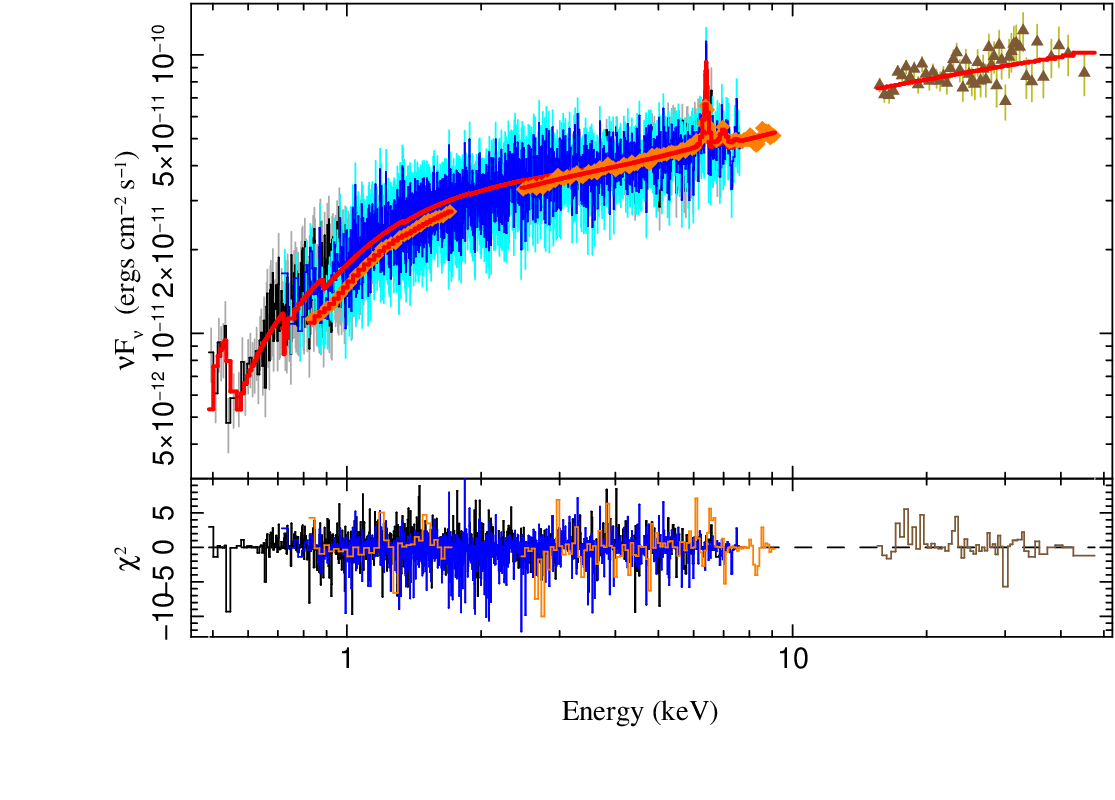}
\includegraphics[width=0.33\textwidth,bb=90 25 597 379]{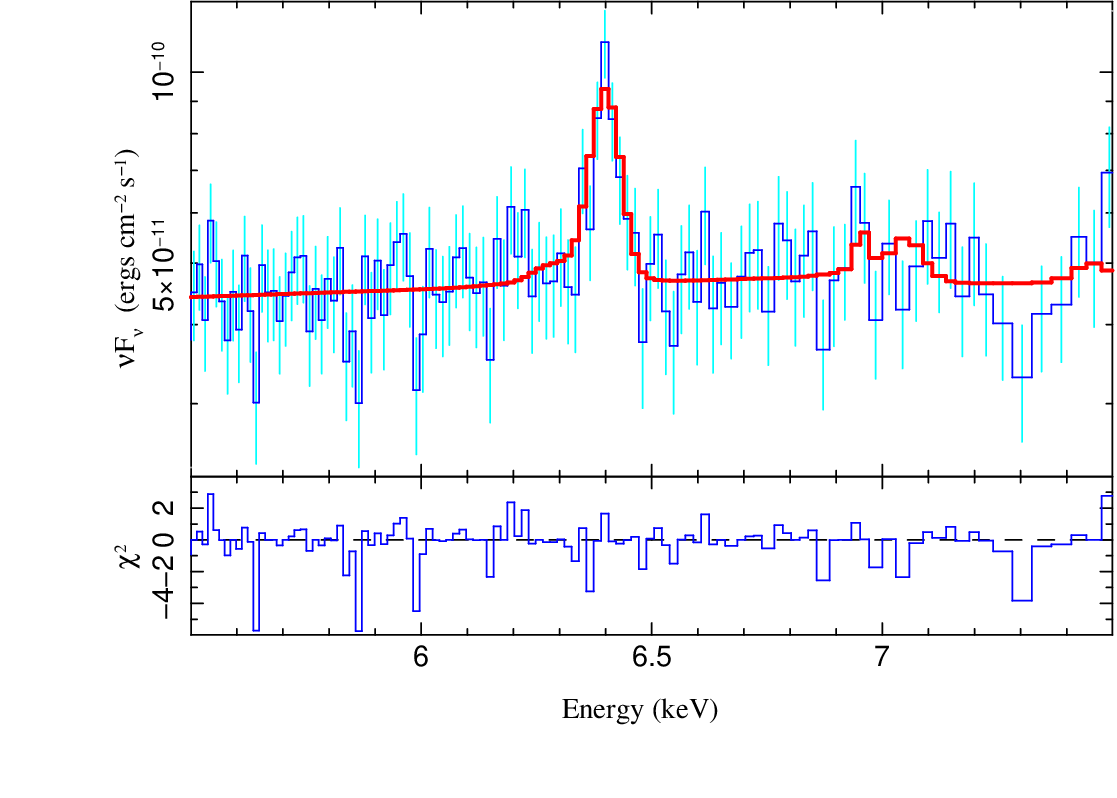}
\includegraphics[width=0.33\textwidth,bb=90 25 597 379]{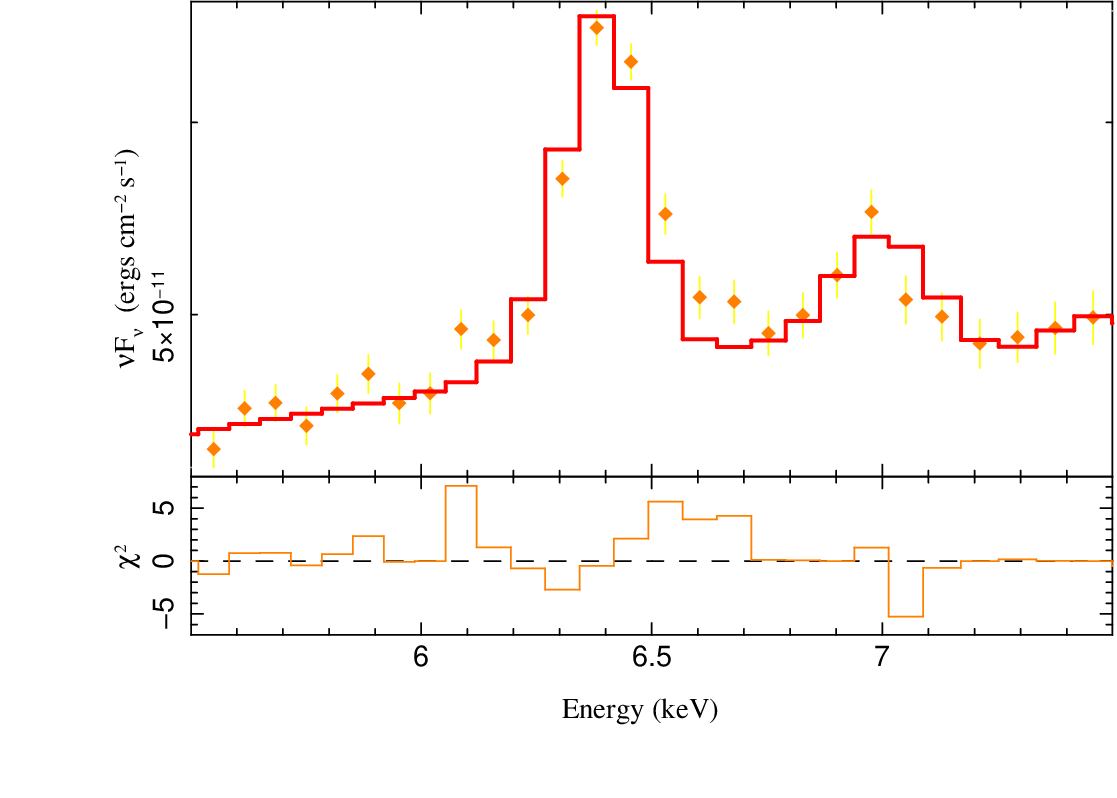}
\end{center}
\caption{Flux-corrected spectra and fit for a model including an absorbed
  power-law continuum, soft excess, \fexxvi line emission, and cold
  Compton reflection (\myt).  {\it Left:} The flux-corrected \meg (black),
  \heg (blue), \xis (orange) and \pin (brown) spectra along with the
  model fit (red). {\it Center:} A close-up of the Fe~K region showing
  only the \heg data.  {\it Right:} A close-up of the Fe~K region
  showing only the \xis data.
\label{fig:mytfit}}
\end{figure*}

\begin{figure*}
\begin{center}
\includegraphics[width=0.33\textwidth,bb=90 25 597 379]{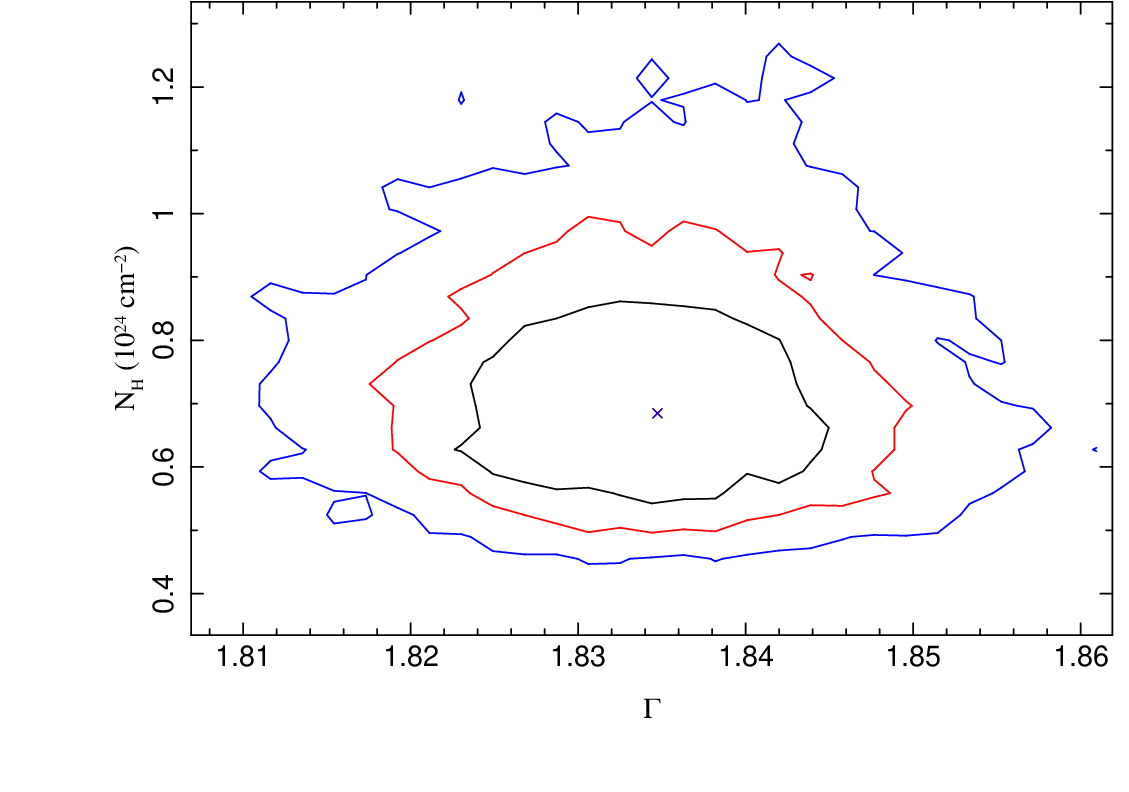}
\includegraphics[width=0.33\textwidth,bb=90 25 597 379]{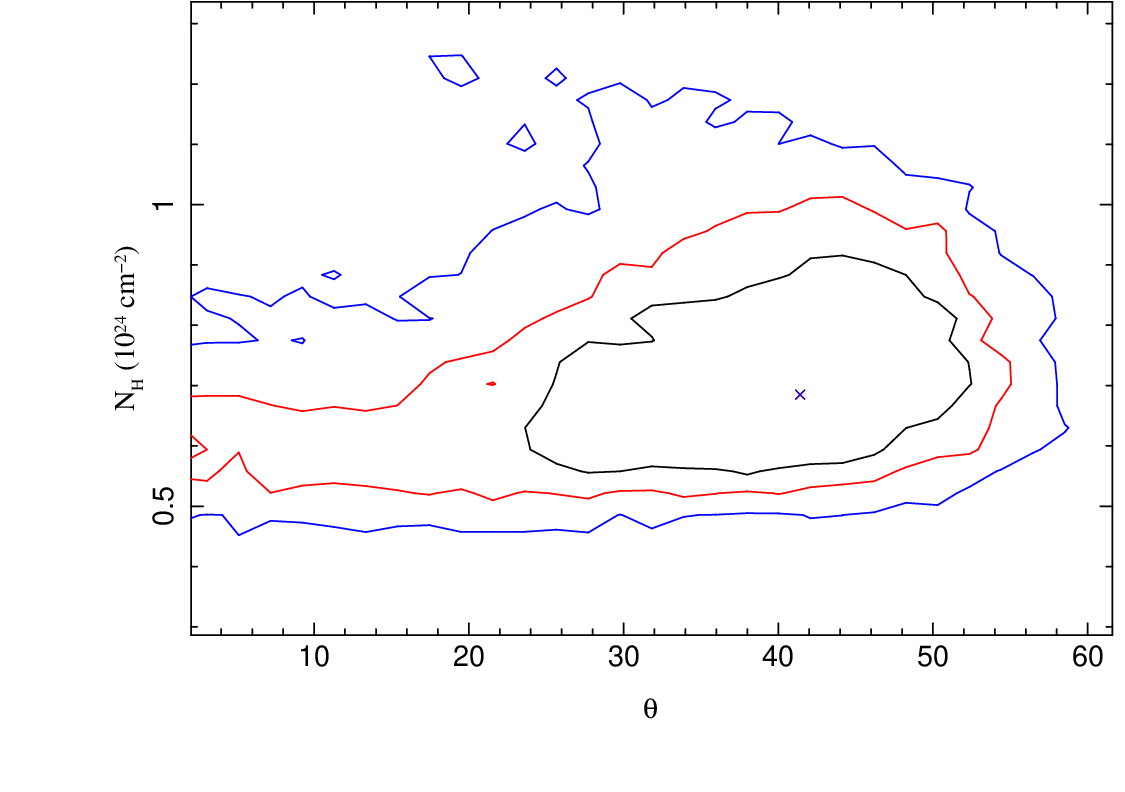}
\includegraphics[width=0.33\textwidth,bb=90 25 597 379]{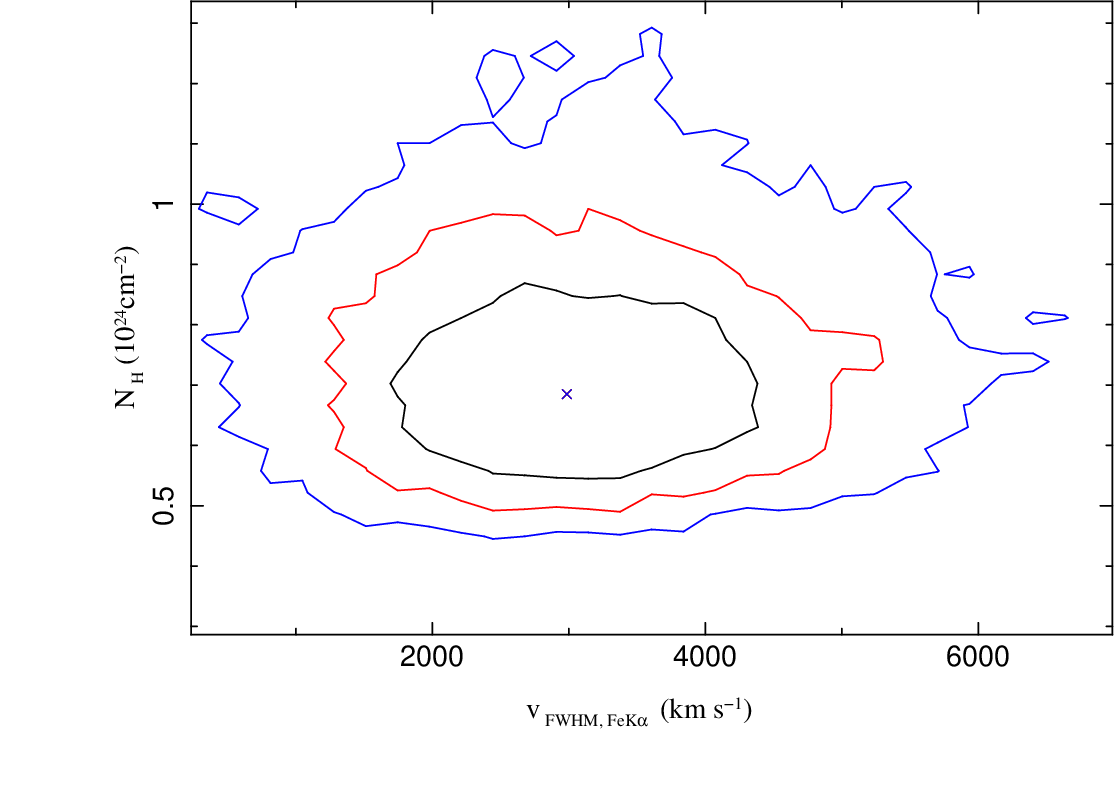}
\end{center}
\caption{The 68\% (black), 90\% (red) and 99\% (blue) confidence 
contours for the model including \myt.  Shown
  are contours for the equatorial column density of the torus versus
  the spectral index of the \hetg data ({\it left}), the
  inclination angle of the reflector ({\it center}), and the FWHM velocity
  width of the \feka emission line.  
\label{fig:mytcont}}
\end{figure*}

In Fig.~\ref{fig:mytfit} we show the model fit to the \heg, \meg,
\xis, and \pin data, including a close-up of the \fek band (center and
bottom panels).  The best-fit parameters are given in
Table~\ref{table:mytpexresults}.  Leaving $\theta$ as a free
parameter, we obtained a reduced $\chi^2$ value of 1.058 for 3288 DoF.
The derived values of the intrinsic power law photon indices,
$\Gamma=1.835^{+0.030}_{-0.012}$ (\chandra) and
$\Gamma=1.743^{+0.006}_{-0.013}$ (\suzaku) are consistent with
published values and those obtained in the \pexmon fit
(\S\ref{sec:pex}).  We obtained an inclination angle of
$\theta=41^{+18}_{-30}$ and an equatorial column density of
\nh$=0.69^{+0.68}_{-0.13}\times10^{24} \ \rm cm^{-2}$ for the torus,
implying that the reflection in the spectrum is due to a possibly
Compton thick torus that is out of the line-of-sight.\footnote{By
  design, lines of sight through the torus would correspond to
  $60^{\circ} \le \theta \le 90^{\circ}$ (MY09).}  Although the
best-fit inclination angle is smaller than the inclination angle found
in the \pexmon fit, we note that this is a model dependent quantity in
both cases.  The toroidal column density and the reflection fraction
obtained from the \pexmon fit are not physically similar parameters
and should not be compared (see, e.g. MY09).  In
Fig.~\ref{fig:mytcont} we show the 68, 90, and 99\% confidence
contours of equatorial column density versus photon index ({\it left})
and inclination angle ({\it center}).  As shown, \nh is well
constrained even though the reprocessing material is out of the line
of sight.

An energy offset for the \myt emission line model table was not
required by the data, so the centroid energies of \feka and \fekb were
intrinsically fixed by the model at 6.404\,keV and 7.058\,keV,
respectively.  Since the \myt model tables do not include kinematics,
we convolved the emission lines with the Gaussian smoothing model
\texttt{gsmooth} and constrained the \feka, \fekb and \nika line
emission to have the same velocity width.  The width of the \feka
emission line was found to be $\sigma=25^{+37}_{-11}$ eV, which
corresponds to a range in full width at half maximum velocity of
1648--7298 km/s.  Assuming Keplerian velocity and a mass of $M_{\rm
  BH}=1.2\times10^{8}M_{\odot}$ \citep{winter:10a}, this places the
reprocessing material at a distance of 0.0129--0.253 pc from the central
black hole.  In Fig.~\ref{fig:mytfit} we show the confidence contours
of \nh versus FWHM velocity width, showing that, with this model, the width is
constrained even to 99\% confidence.

For the thermal emission from the disk, we obtained a peak temperature
of $kT=0.088^{+0.018}_{-0.007}$\,keV, which was assumed to be constant
during both the \chandra and \suzaku observations.  The normalization
of this component was higher for the \chandra data than the \suzaku
data. Both of these results are similar to those obtained in fit
described in \S\ref{sec:pex}.

\section{Conclusions}
We have analyzed the $\sim118$\,ks \chandra-\hetg data together with
$\sim100$\,ks of \suzaku-\xis+\pin data and found the following:
\begin{itemize}

\item The \fexxvi emission line previously reported in \suzaku studies
  of \mcgeight was marginally detected by \chandra.

\item Contrary to previous observations, the \hetg data do not show
  significant evidence of warm absorption.

\item The \meg data reveal evidence of soft excess below the energy
  band probed by \suzaku.  This soft excess is present whether the
  \hetg spectra are fit alone, or in conjunction with the \suzaku
  spectra.

\item The Compton reflection signatures are well-described by material
  that is out of the line of sight.  This is evident from the face-on
  inclination angle that is obtained using the \myt model for the
  reflector and since using the \pexmon model for this
  component did not require additional line-of-sight attenuation
  by cold, high column density material.

\item The reprocessor may be Compton thick, and based upon the
  resolved width of the \feka line is consistent with distances to the
  putative torus.

\item The narrow \feka line emission and associated Compton shoulder
  fits both the \chandra and \suzaku data well.  Although an
  underlying broad component fills in some of the residuals around the
  narrow line in the \xis spectrum, this component is not strongly
  constrained by the data.  It primarily allows a higher Fe abundance
  to be fit for the unsmeared reflector component.

\end{itemize}

The \chandra-\hetg data were crucial for investigating previous
reports of a warm absorber and broad \feka line emission in \mcgeight.
Our analysis leads us to conclude that neither of these components is
required by the data.  Although equally important to our
investigation, \suzaku-\xis+\pin data alone were not capable of fully
characterizing those components, nor the soft excess that was detected
by \chandra below 1\,keV.  Careful modeling of both high spectral
resolution and high signal-to-noise broadband data was required to
show that the Fe~K line emission is well accounted for with cold
Compton reflection (the narrow line emission and associated Compton
shoulder); additional sculpting of the Fe~K region via warm absorption
and/or relativistically-broadened accretion disk reflection is
unnecessary in this case.

\begin{center}

\begin{deluxetable*}{lrr}
\tablecaption{Global Fitting Results \label{table:mytpexresults}}
\tablewidth{0pt}
\tablehead{
\colhead{ Parameter } &  \colhead{{\tt PEXMON} fit} &  \colhead{{\tt MYTorus} fit} }
\startdata
$N_{\rm H, \ los}$ ($10^{22}$ cm$^{-2}$) & 0.386$^{+ 0.021}_{-0.018}$ & 0.347$^{+0.018}_{-0.005}$ \\	
\noalign{\vspace*{0.5mm}}
$kT_{disk}$ (keV)  & 0.091$^{+ 0.007}_{-0.006}$ & 0.088$^{+ 0.018}_{-0.007}$ \\
\noalign{\vspace*{0.5mm}}
$\Gamma_{\rm Chandra}$ & 1.897$^{+0.025}_{-0.025}$ & 1.835$^{+0.030 }_{-0.012}$ \\
\noalign{\vspace*{0.5mm}}
$\Gamma_{\rm Suzaku}$ & 1.818$^{+0.025}_{-0.015}$ & 1.744$^{+0.015}_{-0.006}$ \\
\noalign{\vspace*{0.5mm}}
$N_{\rm PL, \ Chandra} \ (\rm ph \ kev^{-1} \ cm^{-2} \ s^{-1}$ at 1~keV) & 0.0228$^{+0.0045}_{-0.0014}$ & 0.0196$^{+0.0052}_{-0.0012}$  \\
\noalign{\vspace*{0.5mm}}
$N_{\rm PL, \ Suzaku} \ (\rm ph \ kev^{-1} \ cm^{-2} \ s^{-1}$ at 1~keV) & 0.0177$^{+0.0004}_{-0.0004}$ & 0.0170$^{+0.0004}_{-0.0001}$ \\
\noalign{\vspace*{0.5mm}}
$A_{\rm Fe}$ & 0.86$^{+0.23}_{-0.21}$ &  1 {\it f} \\
\noalign{\vspace*{0.5mm}}
$R$ & 3.04$^{+0.43}_{-1.27}$ & \nodata 	\\
\noalign{\vspace*{0.5mm}}
$N_{\rm H, \ torus}$ ($10^{24}$ cm$^{-2}$) & \nodata & 0.685$^{+0.684}_{-0.128}$ \\
\noalign{\vspace*{0.5mm}}
$\theta$ (deg) & $85_{-5.6}$ & 41.4$^{+17.8}_{-30.4}$ \\
\noalign{\vspace*{0.5mm}}
$E_{\rm Fe_{K\alpha}}$ (keV) & 6.4 {\it f} & 6.404 {\it f} \\
\noalign{\vspace*{0.5mm}}
$\sigma_{\rm Fe_{K\alpha}}^{1}$  (eV) & 20$^{+42}_{-13}$ & 25$^{+37}_{-11}$ \\
\noalign{\vspace*{0.5mm}}
$E_{\rm Fe_{K\beta}}$ (keV) & 7.05 {\it f} & 7.058 {\it f}	\\
\noalign{\vspace*{0.5mm}}
$E_{\rm Ni_{K\alpha}}$ (keV) & 7.47 {\it f} & $7.453^{+0.244}_{-0.239}$ \\
\noalign{\vspace*{0.5mm}}
$I_{\rm Ni_{K\alpha}}$ ($10^{-6}$ photons cm$^{-2}$ s$^{-1}$) & \nodata & 3.36$^{+9.41}_{-3.04}$   \\
\noalign{\vspace*{0.5mm}}
$E_{\rm Fe \ XXVI}$ (keV) & 6.951$^{+0.042}_{-0.018}$ & 6.960$^{+0.059}_{-0.054}$\\
\noalign{\vspace*{0.5mm}}
$\sigma_{\rm Fe \ XXVI}$  (eV) &$0^{+53}$ & $0^{+97}$ \\
\noalign{\vspace*{0.5mm}}
$I_{\rm Fe \ XXVI, \ Chandra}$ ($10^{-6}$ photons cm$^{-2}$ s$^{-1}$) & 6.33$^{+8.29}_{-6.33}$	& 4.20$^{+36.7}_{-3.58}$  \\
\noalign{\vspace*{0.5mm}}
$I_{\rm Fe \ XXVI, \ Suzaku}$ ($10^{-6}$ photons cm$^{-2}$ s$^{-1}$) & 9.50$^{+3.55}_{-3.56}$ & 6.87$^{+10.1}_{-3.70}$ \\
\noalign{\vspace*{0.5mm}}
$\chi^2$ (DoF)  & 1.048  (3288) & 1.058 (3288) \\
\noalign{\vspace*{0.5mm}}
$F_{0.5-2 \ \rm keV}$ ($10^{-11} \rm \ erg \ cm^{-2} \ s^{-1}$) & $2.51$ & \nodata \\
\noalign{\vspace*{0.5mm}}
$F_{2-10 \ \rm keV}$ ($10^{-11} \rm \ erg \ cm^{-2} \ s^{-1}$) & $6.57$ & \nodata \\
\noalign{\vspace*{0.5mm}}
$F_{15-50 \ \rm keV}$ ($10^{-10} \rm \ erg \ cm^{-2} \ s^{-1}$) & $1.08$ & \nodata \\

\enddata \tablecomments{Table \ref{table:mytpexresults} contains the
  results of the model fits described in \S\ref{sec:pex} and
  \S\ref{sec:myt}.  0.5--2\,keV and 2--10\,keV flux are for the \heg
  spectra, while the 15--50\,keV flux is for the \suzaku-\pin
  spectrum.  Errors are quoted at the 90\% confidence level for one
  free parameter.  Fixed parameters are denoted with `{\it f}'.}
\tablenotetext{1}{The \feka and \fekb line widths for both models, as
  well as the \nika line width for the pexmon model, were obtained by
  convolving them with {\tt gsmooth}, which gives a width at 6~keV.
  In the \myt fit, the \nika emission line was constrained to have the
  same line width as \feka.  We therefore only quote the \feka line
  width.  The line intensities in these models are self-consistent
  with the other parameters of the models and are therefore not fit
  parameters.  Line energies are in the rest frame of \mcgeight.}
\end{deluxetable*}

\end{center}

\section*{Acknowledgements}
Support for this work was provided by the National Aeronautics and
Space Administration through \chandra Award Number GO1-12147X issued
by the \chandra X-ray Observatory Center, which is operated by the
Smithsonian Astrophysical Observatory for and on behalf of the
National Aeronautics Space Administration under contract NAS8-03060.

\vfill\eject
\goodbreak

\bibliographystyle{jwapjbib}
\bibliography{mnemonic,jw_abbrv,apj_abbrv,bhc,agn,diplom,inst,ns,conferences}

\end{document}